\documentclass[fleqn,12pt]{wlscirep}
\usepackage[utf8]{inputenc}
\usepackage[T1]{fontenc}
\usepackage{setspace} % For double-spacing
\usepackage{lineno}
\usepackage[numbers]{natbib}
\usepackage[cal=cm]{mathalfa}
\usepackage{array}
\usepackage{fancyhdr}

\fancypagestyle{firstpage}{
    \fancyhf{} % Clear all header and footer fields
    \fancyfoot[L]{%
        \makebox[0pt][l]{\rule[2.0ex]{7cm}{0.4pt}} % Short horizontal line (2cm) aligned left
        \hspace{0.5em} % Small space after the line
        \textit{Preprint. Under review.} % Footer text in italics
    }
}

\title{Modeling Multivariable High-resolution 3D Urban Microclimate Using Localized Fourier Neural Operator}

\author[1, 2]{Shaoxiang Qin}
\author[1]{Dongxue Zhan}
\author[1]{Dingyang Geng}
\author[1]{Wenhui Peng}
\author[1]{Geng Tian}
\author[1]{Yurong Shi}
\author[3]{Naiping Gao}
\author[2]{Xue Liu}
\author[1,*]{Liangzhu (Leon) Wang}

\affil[1]{Concordia University, Centre for Zero Energy Building Studies, Department of Building, Civil and Environmental Engineering, Montreal, H3G 1M8, Canada}
\affil[2]{McGill University, School of Computer Science, Montreal, H3A 0E9, Canada}
\affil[3]{Tongji University, School of Mechanical Engineering, Shanghai, 200092, China}

\affil[*]{leon.wang@concordia.ca}

\begin{abstract}

Accurate urban microclimate analysis with wind velocity and temperature is vital for energy-efficient urban planning, supporting carbon reduction, enhancing public health and comfort, and advancing the low-altitude economy. 
However, traditional computational fluid dynamics (CFD) simulations that couple velocity and temperature are computationally expensive. 
Recent machine learning advancements offer promising alternatives for accelerating urban microclimate simulations. 
The Fourier neural operator (FNO) has shown efficiency and accuracy in predicting single-variable velocity magnitudes in urban wind fields. 
Yet, for multivariable high-resolution 3D urban microclimate prediction, FNO faces three key limitations: blurry output quality, high GPU memory demand, and substantial data requirements. 
To address these issues, we propose a novel localized Fourier neural operator (Local-FNO) model that employs local training, geometry encoding, and patch overlapping.
Local-FNO provides accurate predictions for rapidly changing turbulence in urban microclimate over 60 seconds, four times the average turbulence integral time scale, with an average error of 0.35 m/s in velocity and 0.30 $^{\circ}\mathrm{C}$ in temperature. 
It also accurately captures turbulent heat flux represented by the velocity-temperature correlation. 
In a 2 km by 2 km domain, Local-FNO resolves turbulence patterns down to a 10 m resolution. 
It provides high-resolution predictions with 150 million feature dimensions on a single 32 GB GPU at nearly 50 times the speed of a CFD solver. 
Compared to FNO, Local-FNO achieves a 23.9\% reduction in prediction error and a 47.3\% improvement in turbulent fluctuation correlation.

\end{abstract}

\begin{document}
\flushbottom
\maketitle
% \thispagestyle{empty}
\thispagestyle{firstpage}

% \noindent Please note: Abbreviations should be introduced at the first mention in the main text – no abbreviations lists. Suggested structure of main text (not enforced) is provided below.

\nolinenumbers

\section{Introduction}
\label{sec:introduction}
% \textcolor{red}{What's urban microclimate, and why it's important (human comfort, building energy efficiency, low-altitude economy)?} 

% With many countries pursuing carbon neutrality by 2050, there is an unprecedented push to achieve global sustainability goals. 
% The built environment is crucial to this effort, with the building sector responsible for 30\% of global final energy consumption and 26\% of global energy-related greenhouse gas (GHG) emissions \cite{noauthor_buildings_nodate}. 
% Therefore, an accurate representation of the urban microclimate is essential, as it provides city planners and policymakers with insights into the built environment, helping in the development of energy-efficient cities and the reduction of GHG emissions. 
% The urban microclimate refers to localized environmental conditions affected by the built surroundings, with spatial resolutions ranging from 1 mm to 1 km.
% These microclimates differ from rural areas, especially regarding wind speed, wind direction, and air temperature, and affect building energy performance, outdoor thermal comfort, and urban airflow \cite{Sezer-microclimate-review}.
% For example, elevated urban air temperatures, referred to as the urban heat island effect, increase cooling demands and negatively impact the energy performance of the building \cite{HONG2021100871}. 
% Furthermore, as drone delivery and urban air mobility drive new economic opportunities in the emerging low-altitude economy, understanding turbulent urban airflow becomes vital for sustainable development. 

%Tian's version
As the world moves towards ambitious carbon neutrality targets by 2050, the built environment plays a critical role in achieving global sustainability. The building sector, accounting for 30\% of global energy consumption and 26\% of energy-related greenhouse gas (GHG) emissions \cite{noauthor_buildings_nodate}, is a significant factor in this sustainability transition. Accurate modeling and representation of the urban microclimate have become essential, providing city planners and policymakers with data crucial for designing energy-efficient, sustainable urban areas. The urban microclimate encompasses highly localized environmental conditions influenced by the surrounding infrastructure, with spatial resolutions from 1 mm to 1 km. Unlike rural environments, urban microclimates exhibit distinct variations in wind speed, wind direction, and air temperature. These localized climate variations profoundly impact building energy efficiency, outdoor thermal comfort, and airflow dynamics within cities \cite{Sezer-microclimate-review}. For instance, the urban heat island effect—characterized by elevated temperatures in urban areas—amplifies cooling demands, reducing the overall energy efficiency of buildings \cite{HONG2021100871}. Additionally, as innovations like drone delivery and urban air mobility emerge, a comprehensive understanding of urban airflow and turbulence becomes integral to sustainable city planning and development in this new low-altitude economy.

Urban microclimate research primarily utilizes field measurements, experimental methods, and numerical simulations. Field measurements capture temporally varied data such as air temperature, relative humidity, wind speed, and wind direction, offering a realistic view of the daily variations in weather conditions across urban areas \cite{field-measurement-BERNARD2017423, feild-measurement-SUN2024111919, HONG2021100871}. Additionally, advanced tools such as helicopter and satellite thermal imaging aid in gathering temperature distribution data across urban landscapes, thereby enhancing understanding at a broader spatial scale \cite{KLOK_satellites_201223}. Concurrently, wind tunnel experiments—a physical analysis technique extensively used in urban airflow studies—enable the simulation and analysis of air movement under controlled environmental conditions \cite{wind-tunnels-SUN2024111919}.
Despite their contributions, each of these experimental methods has limitations. For example, wind tunnel studies have historically concentrated on airflow, with fewer investigations into the thermal conditions and radiation effects within urban canyons \cite{kanda_passive_2016, marucci_stable_2020, senwen-review-YANG2023110334}. Although some experiments incorporate thermal buoyancy, the emphasis remains on temperature and wind velocity within generic building models or single-street canyon setups. These simplified configurations often omit essential variables, such as solar radiation and shading dynamics, failing to reflect the complex geometries and interactions present in real urban environments \cite{senwen-thesis}.
Furthermore, experimental techniques generally capture data from only a limited number of spatial points, providing a narrower perspective than the comprehensive coverage possible through numerical simulations. This limitation highlights the need for multi-scale approaches that integrate experimental insights with the broader, more detailed outputs achievable in simulation-based studies.

Over recent decades, advancements in computational resources have made numerical simulations, particularly computational fluid dynamics (CFD), increasingly popular for urban microclimate assessment \cite{katal2019modeling, MORTEZAZADEH2022101063, mortezazadeh2019adaptive, mortezazadeh2019slac, mortezazadeh2020solving, tian_turbulence-kinetic-energy_2021}. CFD enables detailed analysis of heat and mass transfer interactions with urban features like buildings and trees, supporting high-resolution simulations that couple velocity and temperature fields. This flexibility allows CFD to model urban microclimates across different scales, from entire cities \cite{tominaga_cfd_2013, MORTEZAZADEH2022101063} to street level \cite{ramponi_cfd_2015,tian_impact_2024}.
However, CFD simulations necessitate a detailed, high-resolution representation of urban geometry, which includes accurately modeling the shapes, sizes, and spatial arrangements of buildings, vegetation, and other urban elements. Precise boundary conditions for all relevant flow variables, such as temperature, wind velocity, humidity, and solar radiation, are also essential to capture the dynamic interactions within urban environments accurately. Additionally, CFD simulations demand substantial computational resources to handle the high volumes of data and complex calculations required to resolve fine-scale processes across large domains. These requirements make CFD a resource-intensive approach, particularly for large urban-scale simulations that may require advanced computational infrastructure and optimization techniques to manage the extensive computational load effectively \cite{barlow_progress_2014,blocken_50_2014}.

With the recent progress in artificial intelligence, deep learning models, particularly deep neural networks, have demonstrated potential as more efficient surrogate models for solving partial differential equations (PDEs) than traditional numerical methods. 
After being trained on PDE equations or solution data, neural networks can quickly predict PDE solutions for unseen cases. 
The two main deep learning paradigms for solving PDEs are learning physics-informed models and purely data-driven models. 
Physics-informed models, like physics-informed neural networks (PINNs) \cite{raissi2019physics, pinn}, use physics-based loss functions and optional data observations to guide their training. 
This physics loss introduces knowledge of the underlying PDEs, helping to produce more accurate predictions in data-limited or even data-free scenarios. 
Like traditional numerical solvers, physics-informed models rely on the governing PDE equations. 
In contrast, pure data-driven models do not require any physics knowledge and can be trained solely on historical or synthetic data \cite{azizzadenesheli2024neural, deeponet, hao2023gnot}. 
As a trade-off, they tend to require a larger amount of training data compared to physics-informed models. 
%One significant advantage of pure data-driven models is their ability to overcome the inaccuracies of human-developed physical models when dealing with highly complex natural dynamics. 
One significant advantage of pure data-driven models is their ability to overcome the limitations of traditional physical models, which often rely on simplified assumptions that may not fully capture the complexities of real-world natural dynamics.
Recent advanced deep learning weather models, such as Pangu \cite{pangu} and GraphCast \cite{graphcast}, can be trained end-to-end on weather data and have demonstrated the ability to produce more accurate predictions than the leading numerical weather models. NeuralGCM \cite{neuralgcm} further improves forecasting performance through the integration of machine learning components and a differentiable solver.

% \textcolor{red}{Introduce neural operators for solving PDEs.}
Common deep learning models, such as convolutional neural networks (CNNs) \cite{cnn}, graph neural networks (GNNs) \cite{gnn}, and transformers \cite{vit}, are designed for spatially or temporally discrete data. 
These models are trained on fixed-resolution datasets and cannot generalize to other spatial or temporal points. 
This restriction poses challenges in modeling continuous natural processes like fluid dynamics. 
To address this, neural operators \cite{azizzadenesheli2024neural, deeponet} have been developed as resolution-invariant models for scientific machine learning. 
Neural operators approximate operators—mappings between continuous functions—and can make predictions across different resolutions, even when trained on fixed-resolution data.
Among the neural operators, Fourier neural operator (FNO) \cite{fno} demonstrates superior accuracy and efficiency and have been applied to a range of fields, such as modeling weather \cite{fourcastnet}, fluid turbulence \cite{li2023long}, materials \cite{rashid2022learning}, smectic waves \cite{li2023solving}, and plasma \cite{gopakumar2024plasma}. 
Unlike traditional neural networks like CNN, which are parameterized in the physical domain, most of the parameters in FNO are defined in the Fourier domain. 
This enables FNO to learn resolution-invariant representations from the data. Additionally, the Fourier transformation in FNO functions similarly to a reduced-order model, as turbulence data typically concentrates energy in low-frequency modes \cite{kolmogorov1991local}. 
FNO commonly truncates high-frequency modes to improve memory and time efficiency during training and predicting.
% with minimal compromise to accuracy on less complicated PDEs.
Recently, FNO has been demonstrated to be an effective approach for modeling velocity magnitude field in urban microclimate \cite{peng2024fourier}.
%The Fourier neural operator represents a breakthrough in machine learning applied to fluid dynamics, specifically in modeling turbulent flows with zero-shot super-resolution capabilities. This method leverages the Fourier transform and neural networks to achieve significant advancements in accurately simulating and predicting turbulent fluid behaviors. This approach not only enhances predictive accuracy but also enables finer resolution of turbulent flow phenomena without requiring explicit training data at those resolutions. This capability is particularly significant in fluid dynamics where capturing complex turbulent structures accurately and efficiently has been a long-standing challenge.

% \textcolor{red}{FNO's flaw in capturing fine details in the flow field.} 
This work focuses on making fast and accurate predictions for multivariable high-resolution 3D urban microclimate, specifically, the three directional wind velocity and temperature. 
Previous research is limited to predicting the velocity magnitude (one variable) in urban wind fields \cite{peng2024fourier}. 
When applied to multivariable urban microclimate prediction in complex urban environments, vanilla FNO faces three critical limitations: blurry output quality, extensive GPU memory usage, and substantial data demands. 
First, FNO’s predictions often appear blurry because urban microclimate with temperature differences involves complex physical processes, resulting in intricate small-scale features in the flow field. 
The truncation of high-frequency modes in each Fourier layer limits FNO’s capacity to capture these small-scale features accurately. 
Second, FNO’s GPU memory usage during prediction scales nearly proportionally with the total grid number of the wind field data. 
Given the high dimensionality of multivariable 3D data, FNO often requires excessive memory that may exceed available resources. 
Third, the curse of dimensionality poses a significant challenge for modeling complex physical processes with machine learning methods. 
In multivariable 3D microclimate simulations, CFD data dimensions—including three directional wind velocity and temperature across all grids—can reach hundreds of millions, while only thousands of time steps may be available for training. 
The dataset becomes extremely sparse in such a high-dimensional space, making the machine learning model easily overfit and perform poor generalization.
To overcome these limitations, we introduce the localized Fourier neural operator (Local-FNO), a novel approach for learning and predicting multivariable high-resolution 3D urban microclimate. 
The core design is to train the FNO on smaller, local patches of the domain rather than applying it to the entire domain at once. 
This local training strategy addresses all three limitations. 
First, Local-FNO effectively captures small-scale features in the wind field. 
With the same number of Fourier modes, the smaller domain size produces shorter Fourier wavelengths, allowing Local-FNO to represent finer details than FNO. 
Second, Local-FNO makes independent predictions for each small patch at every time step, reducing peak memory usage. 
This approach also allows for parallel predicting when multiple GPUs are available. 
Third, by dividing the entire domain into smaller patches and training a shared Local-FNO on each, we reduce the data dimensionality and increase the number of training samples, thus lowering the data requirements and enhancing FNO’s generalization ability.
Two methods are further incorporated into Local-FNO to enhance its performance and counteract the side effects of local training: geometry encoding and patch overlapping. 
Geometry encoding informs Local-FNO with local building geometries using signed distance functions. 
Patch overlapping allows wind flow to interact across neighboring patches, minimizing discontinuities at the patch boundaries.
With these designs, Local-FNO is capable of producing accurate and fast predictions for multivariable, high-resolution 3D urban microclimate. Its effectiveness and efficiency are validated through comprehensive evaluations.

Several prior studies have aimed to enhance neural operators' ability to capture small-scale features and sharpen predictions beyond FNO. 
U-FNO \cite{wen2022u} and IU-FNO \cite{li2023long} incorporate a U-Net structure into FNO. U-Net \cite{ronneberger2015u} employs a U-shaped architecture with downsampling and upsampling to capture multiscale features. 
HANO \cite{liu2024mitigating} decomposes the input-output mapping into hierarchical levels and applies self-attention across these levels to improve multiscale feature capture. 
LNO \cite{ye2024locality} combines convolution in physical space with linear transformations in Legendre spectral space to learn local operators from PDE data. 
However, unlike FNO, these models \cite{wen2022u, li2023long, liu2024mitigating, ye2024locality} lack resolution invariance due to their reliance on convolutions or attentions within discrete spatial domains. 
Some methods can retain FNO's resolution-invariant property. 
For instance, a local integral kernel \cite{Liu-SchiaffiniB24} is introduced to replace discrete convolution, achieving resolution-invariant local convolution by applying a continuous convolutional kernel approximated by summing over a fixed set of triangular functions on discrete data points. 
Compared to this local integral kernel, our proposed Local-FNO demonstrates that Fourier modes defined over local regions are also effective at capturing resolution-invariant local features with a more straightforward implementation.
Other methods include PDE-Refiner \cite{lippe2024pde}, which uses a denoising diffusion model to refine predictions from various machine learning models, enhancing their performance on non-dominant frequencies, although it is less effective for FNO than other models. 
SpecB-FNO \cite{qin2024toward} enhances FNO's prediction of non-dominant frequencies by sequentially training multiple FNOs, with each learning to predict the residual of its predecessor. 
However, its improvement is limited when FNO uses relatively few Fourier modes.
All methods mentioned above, except for LNO, require training on the entire domain of PDE data, making them unsuitable for the extremely high-dimensional data in this study given limited computational resources.

\section{Methodology}
\label{sec:methodology}

\subsection{Numerical simulations}
% \textcolor{red}{Introduce CityFFD. Describe the algorithm of CityFFD and how it is different from other common solvers.} 
The airflow and temperature data generated by City Fast Fluid Dynamics (CityFFD) serve as training and testing data for FNO and Local-FNO models. By employing a semi-Lagrangian approach and fractional stepping method with various novel numerical schemes, CityFFD enhances model accuracy and reduces computational costs \cite{mortezazadeh2017high,mortezazadeh2019adaptive}. A large eddy simulation (LES) model is implemented in CityFFD to capture turbulence within the atmospheric boundary layer \cite{mortezazadeh2020solving}. CityFFD integrates a fourth-order numerical interpolation scheme to minimize numerical dissipation and dispersion errors, even on coarse grids. Developed with CUDA-C++ for GPU processing, CityFFD is designed to predict local microclimate features and model large-scale urban aerodynamics. Recent studies have shown that CityFFD can effectively model urban-scale building scenarios, with its fundamental theory detailed in prior research \cite{MORTEZAZADEH2022101063, mortezazadeh2019adaptive}. Previous studies have applied CityFFD in urban areas to evaluate wind and thermal scenarios. CityFFD can also integrate with the Weather Research and Forecasting Model (WRF) and couple with building energy models \cite{CityFFD+EnergyPlus, luo2022data}. A series of validation cases, comparing CityFFD results with wind tunnel and real urban data, have been conducted under both isothermal and non-isothermal conditions. These validations demonstrate that CityFFD performs well in predicting airflow and air temperature.

CityFFD sovles the continuity, momentum, and energy equations presented as:
\begin{equation}
\label{cityffd-equation}
\begin{aligned}
    \nabla \cdot \vec{U} &= 0 \\
    \frac{\partial \vec{U}}{\partial t} + (\vec{U} \cdot \nabla) \vec{U} &= -\nabla p 
    + \left(\frac{1}{Re} + \nu_t\right) \nabla^2 \vec{U} - \frac{Gr}{Re^2} \, \theta \\
    \frac{\partial \theta}{\partial t} + (\vec{U} \cdot \nabla) \theta &= 
    \left(\frac{1}{Re \cdot Pr} + \alpha_t\right) \nabla^2 \theta \ ,
\end{aligned}
\end{equation}
where $\vec{U}$, $\theta$, and $p$, $Re$, $Gr$, $Pr$, $v_t$, and $\alpha_t$ represent the velocity vector, temperature, pressure, Reynolds number, Grashof number, Prandtl number, turbulent viscosity, and turbulent thermal diffusivity, respectively.

The advection terms in Equation \ref{cityffd-equation} are solved using a Lagrangian approach, which calculates the air temperature and wind velocity at position $S_c^{n+1}$ based on the values of $\vec{U}$ and $\theta$ at position $S_c^n$, as shown in Equation \ref{cityffd-equation-lagrangian}:
\begin{equation}
\label{cityffd-equation-lagrangian}
    S_c = \vec{U} \mathrm{d} t \rightarrow S_c^n \approx S_c^{n+1} - \vec{U}\Delta t \ .
\end{equation}
The turbulent viscosity is calculated by the following equation:
\begin{equation}
\label{eq:turbulence}
v_t = (c_s \Delta)^2 |S| \ ,
\end{equation}
 where $c_s$ is the Smagorinsky constant, typically ranging from 0.1 to 0.24, $\Delta$ is the filter width, and $S$ is the large-scale strain rate. CityFFD employs a fourth-order interpolation scheme to model airflow on coarse grids and reduce high dissipation errors. Detailed methodology can be found in earlier work \cite{mortezazadeh-thesis}.

\subsection{Data description}
% \textcolor{red}{Describe the Montreal case: physical domain size, wind setting, grid size, time interval... Also, the validation on CFD results.}

\begin{figure} [htbp]
    \centering
    \includegraphics[width=0.8\textwidth]{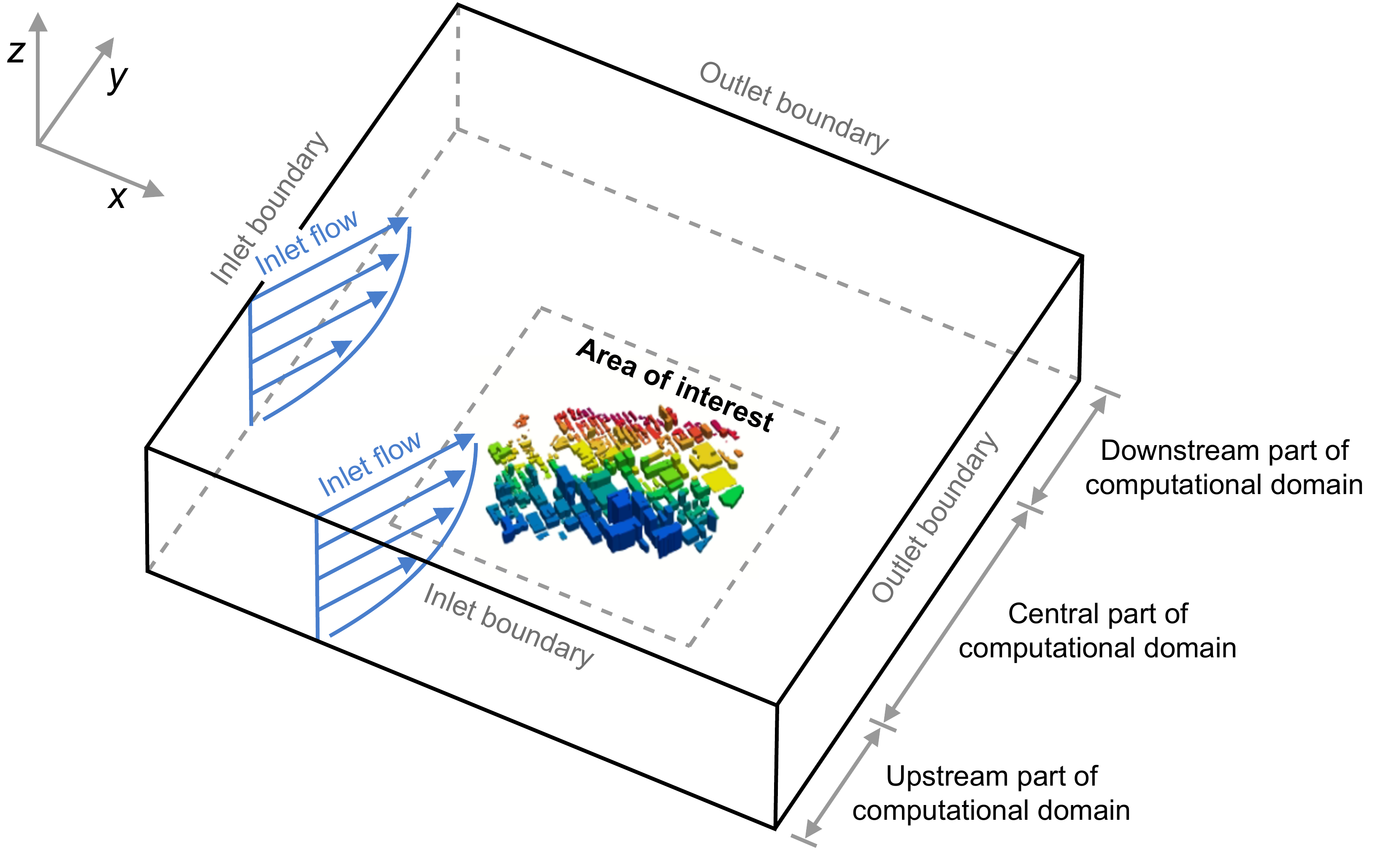}
    \caption{Computational domain for CFD simulation}
    \label{fig:domain}
\end{figure}

The studied building cluster is located in downtown Montreal, covering a $2 \text{ km} \times 2 \text{ km}$ area with 274 buildings, the tallest of which is 200 m. Buildings are densely distributed in the central $1.5 \text{ km} \times 1.5  \text{ km}$ section. 
The total computational domain is $4 \text{ km} \times 4 \text{ km}$ with a height of 1 km, as demonstrated in Figure \ref{fig:domain}. It consists of 54 million grids, with the minimum grid size near buildings being $4 \text{ m} \times 4 \text{ m}$ horizontally and 1 m vertically. 
Vertical boundaries of the domain function as inlets or outlets, depending on wind direction, while the ground boundary is set as a wall and the top as symmetry. Building surface temperatures, calculated by the City Building Energy Model (CityBEM) \cite{katal2019modeling}, serve as boundary conditions in CityFFD simulations.
Inlet boundary conditions were taken from Dorval airport weather station, located 13 km from the studied area. Simulations were conducted for July 16, 2013, at 17:00, with an air temperature of 30.7 $^{\circ}\mathrm{C}$, wind speed of 4.16 m/s, and wind direction of 23$^{\circ}$ from $y$-axis toward $x$-axis. The time step was set at 0.2 s.

% The studied building cluster is situated in downtown Montreal with the size of 2 km × 2 km, containing 274 buildings with a maximum of 100 m height (the central 1.5 km × 1.5 km portion is densely developed). The overall computational domain size is 4 km × 4 km with 1 km height. Entire domain consists of a total 54 million cells, with the minimum mesh size near the building is 4 m × 4 m with 1 m height. The vertical surfaces of the computational domain serves as inlet or outlet, depending on the wind direction. The ground boundary condition was treated as a wall, and the top boundary condition was symmetry. Building surface temperature is calculated by city building energy model (CityBEM) \cite{katal2019modeling} and used as a boundary condition in CityFFD simulations. The inlet boundary conditions were collected from Dorval airport weather station, about 13 km away from the studied area. The simulation was conducted for 7/16/2013 17:00, with an air temperature of 30.7 $^{\circ}\mathrm{C}$, wind speed is 4.16 m/s, wind direction of 23 $^{\circ}$. The time step was set as 0.2 s.

Validation for this case has been studied in previous literature \cite{senwen-thesis}, showing an RMSE of 0.97 $^{\circ}\mathrm{C}$ for air temperature and 0.257 m/s for wind velocity. The validation was conducted over two consecutive days during a heatwave, using onsite measurements documented by Environment and Climate Change Canada as part of a temporary field campaign in the summer of 2013.

\subsection{Neural operators and Fourier neural operator}
Neural operators are designed to learn the mapping between input and output functions defined in continuous spaces. Below, we introduce the problem settings for neural operators as established in prior research \cite{fno, kovachki2021universal}. Let $D \subset \mathbb{R}^d$ be a bounded domain with spatial dimension $d \in \mathbb{N}$, where, for example, $d=3$ represents a 3D space. Consider $a \in \mathcal{A}(D; \mathbb{R}^{d_a})$ as the input function $a : D \to \mathbb{R}^{d_a}$, which has $d_a \in \mathbb{N}$ components, and $u \in \mathcal{U}(D; \mathbb{R}^{d_u})$ as the output function $u : D \to \mathbb{R}^{d_u}$, which has $d_u \in \mathbb{N}$ components. $d_a$ and $d_u$ can be considered as the variable dimensions defined on the input space and output space, respectively. Both $\mathcal{A}(D; \mathbb{R}^{d_a})$ and $\mathcal{U}(D; \mathbb{R}^{d_u})$ are  Banach spaces, and let $\mathcal{G}^\dagger : \mathcal{A}(D; \mathbb{R}^{d_a}) \to \mathcal{U}(D; \mathbb{R}^{d_u})$ denote the true operator between these spaces. The goal of operator learning is to approximate the true operator $\mathcal{G}^\dagger$ by learning a model $\mathcal{G}_\theta$, parameterized by $\theta$, using a finite set of discretized input-output samples from $\mathcal{G}^\dagger$. The neural operator proposed by Li et al. \cite{li2020neural} is defined as: 
\begin{equation}
\mathcal{G}_\theta = \mathcal{Q} \circ \mathcal{L}_L \circ \mathcal{L}_{L-1} \circ \cdots \circ \mathcal{L}_1 \circ \mathcal{P} \ ,
\end{equation}
where $L \in \mathbb{N}$ is the depth of the model. The mapping begins with the lifting operator $\mathcal{Q}: \mathcal{A}(D; \mathbb{R}^{d_a}) \to \mathcal{U}(D; \mathbb{R}^{d_v})$, which raises the input channels from $d_a$ to $d_v$. The number of channels remains $d_v$ through all intermediate layers and is projected back to the output channels using the projection operator $\mathcal{P}: \mathcal{U}(D; \mathbb{R}^{d_v}) \to \mathcal{U}(D; \mathbb{R}^{d_u})$. Both $\mathcal{Q}$ and $\mathcal{P}$ are pointwise operators, meaning they act independently at each point in the $d$-dimensional space. $\mathcal{L}_1, \mathcal{L}_2, \dots, \mathcal{L}_L$ represent nonlinear operator layers, which are key to achieving the mapping between continuous spaces. For $\ell \in L$, the mapping $\mathcal{L}_\ell : \mathcal{U}(D; \mathbb{R}^{d_v}) \to \mathcal{U}(D; \mathbb{R}^{d_v})$ keeps the channels $d_v$.

In this work, the nonlinear operator layer is defined as follows: 
\begin{equation}
\mathcal{L}_\ell(v)(x) = \sigma\left( \mathcal{M}_\ell\left( \left(\mathcal{K}(\theta_\ell)v\right)(x)\right) + W_\ell v(x) + b_\ell(x) + v(x) \right) \ ,
\label{eq:layer}
\end{equation} 
where $x \in \mathbb{R}^d$ is the spatial coordinate and $v \in \mathcal{U}(D; \mathbb{R}^{d_v})$ is the function in the intermediate layers. $\mathcal{L}_\ell$ consists of a kernel operator $\left(\mathcal{K}(\theta_\ell)v\right)(x)$, a pointwise nonlinear feedforward network $\mathcal{M}_\ell$, a pointwise affine mapping $W_\ell v(x) + b_\ell(x)$, an identity mapping $v(x)$, and a nonlinear activation function $\sigma$. 
In contrast to its original definition \cite{li2020neural}, we incorporate two design modifications frequently used in the latest research: identity mappings and point-wise feedforward neural networks (FFNs). 
Identity mappings \cite{he2016deep} enable deep neural networks to scale a large number of layers and have demonstrated effectiveness in neural operators \cite{ffno}. 
The point-wise FFN, inspired by the transformer's architecture \cite{vaswani2017attention}, enhances the information interaction between channels.

Fourier neural operator \cite{fno} is a specialized type of neural operator with advantageous performance in solving PDEs. The kernel operator in FNO is defined as a complex linear transformation in Fourier space, specifically: 
\begin{equation}
(\mathcal{K}(\theta)v)(x) = \mathcal{F}^{-1} \left( \mathcal{R}_\theta(k) \cdot \mathcal{F}(v)(k) \right)(x) \ .
\end{equation}
Here, $\mathcal{F}$ and $\mathcal{F}^{-1}$ represent the Fourier transform and inverse Fourier transform, respectively. $k \in \mathbb{Z}^d$ is the complex index in Fourier space, and $\mathcal{R}_\theta(k) \in \mathbb{C}^{d_v \times d_v}$ is a complex matrix that performs linear mapping in this space. In FNO, only the kernel operator $\mathcal{K}$ is not a pointwise operator. The kernel operator enables information interaction across all spatial locations and acts as a continuous global convolution kernel.

In practice, FNO is trained on samples defined on discretized grids over $D$. The Fourier transform $\mathcal{F}$ and inverse $\mathcal{F}^{-1}$ are computed using the fast Fourier transform (FFT) algorithm \cite{fft}. Since FFT operates only on structured grids, FNO can only use data that is either defined on or can be converted to a structured grid \cite{geo-fno}. Pointwise mappings $\mathcal{Q}$, $\mathcal{P}$, and $\mathcal{M}$ are typically implemented using a multilayer perceptron (MLP) \cite{mlp} with one hidden layer. These mappings operate independently on each grid point.

For inputs defined on structured grids, the transformed inputs in the Fourier domain retain the same structured grids with an identical number of grid points. Each grid index $k$ in the Fourier domain, referred to as a mode, corresponds to a sinusoidal wave in the spatial domain with a specific frequency and direction. Performing complex linear transformations on all frequencies is computationally expensive, so FNO truncates high-frequency components across all spatial dimensions. Given the spatial resolution $s_1 \times \cdots \times s_d$ and truncation frequencies $k_{\text{max},j}$, $j = 1, \dots, d$, the set of preserved modes after truncation is denoted as:
\begin{equation}
Z_{k_{\text{max}}} = \left\{ (k_1, \dots, k_d) \in \mathbb{Z}^{s_1} \times \cdots \times \mathbb{Z}^{s_d} \; \middle| \; k_j \leq k_{\text{max},j} \text{ or } s_j - k_j \leq k_{\text{max},j}, \; \text{for} \; j = 1, \dots, d \right\} \ .
\end{equation}
After truncation, the total number of complex parameters defined in $\mathcal{R}_\theta$ is proportional to $k_{\text{max},1} \times \cdots \times k_{\text{max},d} \times d_v \times d_v $. 
% For spatial dimension $j$, the total number of remaining modes is $2k_{\text{max},j}$. Due to the centrosymmetry of the Fourier transform for real-valued functions, we only need to define parameters for half of the total modes. 
For PDE data of lower complexity, FNO can still perform well after truncating high frequencies due to the Fourier transform's efficient compression property. In data sampled from continuous space or time, such as natural images \cite{ruderman1993statistics} and turbulent flows \cite{kolmogorov1991local}, the majority of the energy is concentrated in lower frequencies. 

\subsection{Localized Fourier Neural Operator}
\label{sec:local-fno}
In this work, we propose the Localized Fourier Neural Operator to address three key challenges in applying FNO to multivariable 3D urban microclimate prediction: blurry output quality, extensive GPU memory usage, and substantial data demands. 
Local-FNO has three major design components: a local training strategy, geometry encoding, and patch overlapping. 
The local training strategy addresses all three main limitations, while geometry encoding and patch overlapping mitigate its side effects to further enhance Local-FNO’s performance. 
The model architecture of Local-FNO builds upon FNO while integrating recent improvements from neural operator research, as discussed in equation \ref{eq:layer}.

\begin{figure} [htbp]
    \centering
    \includegraphics[width=1.0\textwidth]{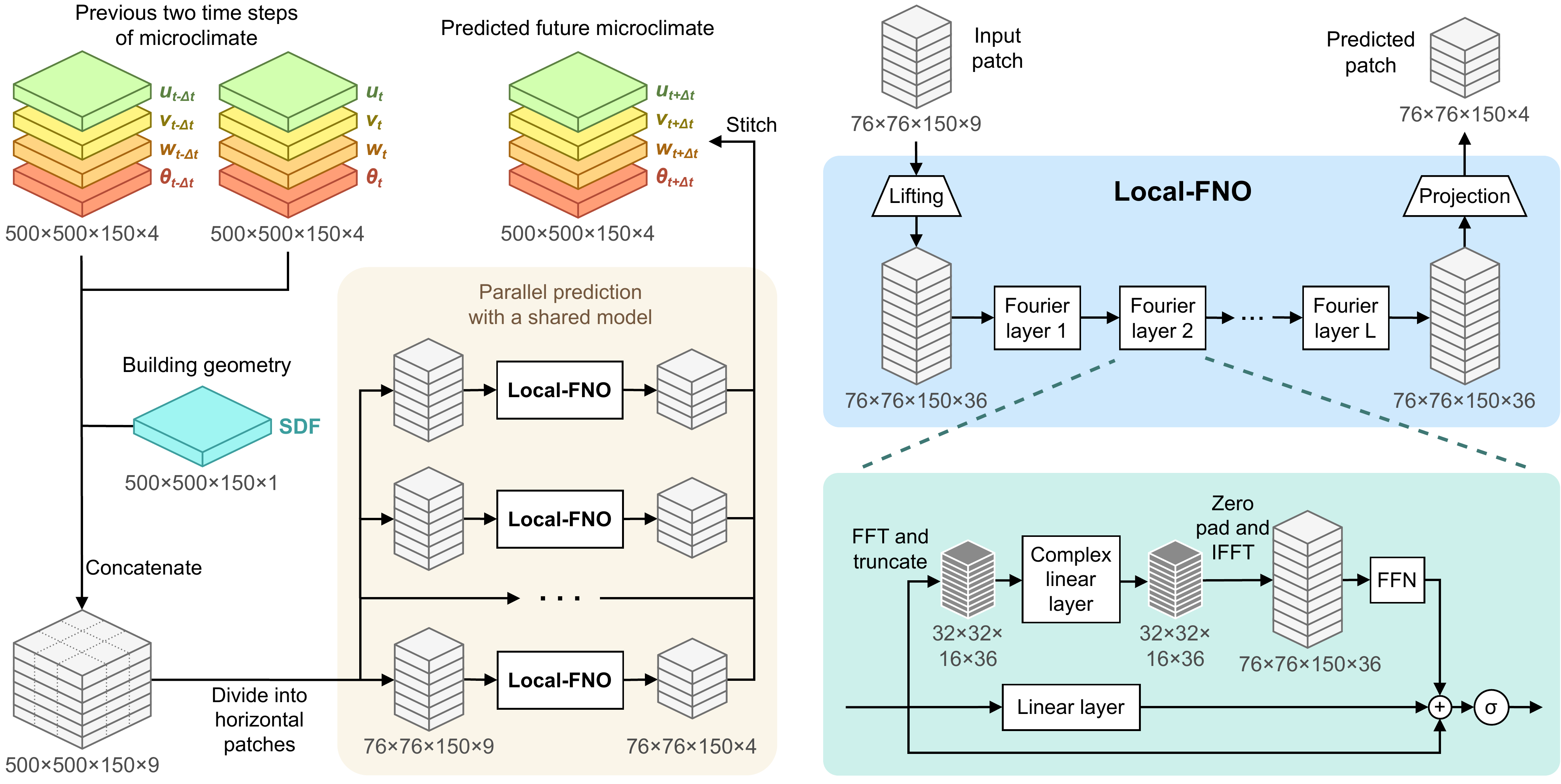}
    \caption{Overview of Local-FNO design. Left: prediction procedure of Local-FNO. Right: the architecture of Local-FNO and its Fourier layers.}
    \label{fig:local-fno}
\end{figure}

As shown in Figure \ref{fig:local-fno}, Local-FNO predicts future microclimate within each local patch. 
In a 3D computational domain, patches are divided along two horizontal axes only, as microclimate complexity is typically lower along the vertical axis than the horizontal axes. 
The patches are divided in a uniform grid structure, aligned across both horizontal axes, with neighboring patches overlapping to share boundary areas. 
The total number of patches and the overlap rate can be adjusted for specific tasks. 
Figure \ref{fig:local-fno} illustrates an example with $8 \times 8 = 64$ patches and a 20\% overlap, which we apply in our prediction task.

Local-FNO is designed to predict urban microclimate variables at a future time step $t + \Delta t_{\text{pred}}$, based on data from two prior time steps, $t - \Delta t_{\text{pred}}$ and $t$, along with the signed distance function (SDF) \cite{osher1988fronts} of the building geometry. 
Unlike in CFD, where the time step $\Delta t$ is restricted by CFL conditions \cite{cfl}, machine learning methods do not impose such rigid criteria. 
As a result, the prediction interval $\Delta t_{\text{pred}}$ in machine learning can be considerably larger than the time step $\Delta t$ used in CFD simulations. 
Using two prior time intervals as inputs captures the dynamics of urban microclimate variables effectively—a strategy also employed in Graphcast \cite{graphcast}. 
Since urban microclimate is strongly influenced by local building geometry, we encode each local patch’s geometry using its SDF. 
This geometry encoding provides Local-FNO with the necessary building information to improve wind field predictions. 
Although Local-FNO is trained to predict only a single time step of microclimate ahead, it can be applied iteratively for rollout predictions by feeding its previous prediction as input for the next step. 

\begin{figure} [htbp]
    \centering
    \includegraphics[width=0.4\textwidth]{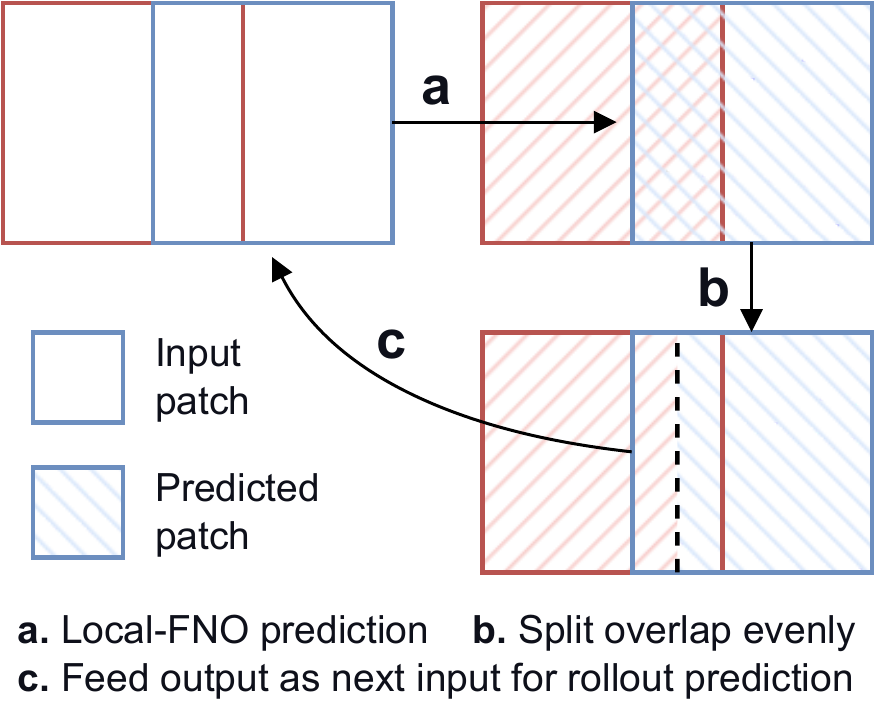}
    \caption{Patch overlapping in Local-FNO. Patch overlapping enables information exchange between patches and avoids using Local-FNO’s own boundary predictions.}
    \label{fig:rollout}
\end{figure}

Patch overlapping is an essential design for Local-FNO to ensure smoother transitions between patches. 
Simply predicting each patch independently can cause severe discontinuity between them for two main reasons.
First, each patch lacks information about the wind flow in neighboring patches, preventing continuous movement across patch boundaries. 
Second, FNO is designed for cyclical boundary conditions since it operates in the Fourier space. 
When actual boundaries are not cyclical, FNO's prediction errors are often more pronounced near patch boundaries.
Figure \ref{fig:rollout} illustrates how the patch overlapping approach is used to address these issues. 
Each input patch partially overlaps with its neighboring patches. 
First, Local-FNO takes each input patch, including its overlapping areas, to predict the corresponding output patch. 
In the output, overlapping regions are split evenly between neighboring patches, meaning that each patch’s boundary area is predicted by neighboring patches. 
This design enables information exchange between patches and avoids using Local-FNO’s own boundary predictions, which tend to have higher errors. 
The final output of each patch is then used iteratively for rollout predictions.

By training and predicting in smaller patches, Local-FNO effectively addresses three key limitations of FNO. 
First, Local-FNO produces much sharper and more accurate predictions than FNO. 
This improvement arises because, with the same number of Fourier modes, a smaller domain size generates shorter Fourier wavelengths. 
For instance, in a one-dimensional domain of 1000 m, the first four Fourier wavelengths are 1000 m, 500 m, 333.3 m, and 250 m. 
When applied to a 100 m domain, these wavelengths shorten to 100 m, 50 m, 33.3 m, and 25 m. 
In FNO, shorter wavelengths allow for better capture of small-scale features in the urban wind field.
Secondly, Local-FNO breaks down the full-domain microclimate prediction into smaller, independent tasks, greatly lowering memory demands. 
This design also supports scalability for large-scale microclimate predictions, as the workload can be distributed across multiple GPUs.
Third, Local-FNO improves generalization and requires less data to make accurate predictions. 
By focusing on local features, it learns from smaller, more relevant wind patterns, reducing the risk of capturing spurious global correlations in data-limited scenarios. 
Dividing data into smaller patches also increases data efficiency, as it reduces dimensionality and expands the number of training samples without requiring extra raw data.

\subsection{Model training}
For each time step, the CityFFD simulation has a total grid size of $600 \times 600 \times 150$ along the $x$, $y$, and $z$ axes, covering a physical area of $4\text{ km} \times 4 \text{ km} \times 1 \text{ km}$. The central area of interest has a grid size of $500 \times 500 \times 150$, corresponding to a physical area of $2\text{ km} \times 2 \text{ km} \times 1 \text{ km}$. Only data from this area of interest is used for machine learning. We conduct a CityFFD simulation for a total of 23000 steps, with a time step $\Delta t$ of 0.2 s. The flow field reaches full development around 8000 steps. For the remaining 15000 steps, we save data at every 10th step, resulting in 1500 saved time steps that represent 50 minutes of wind field data. Each time step contains three directional wind speeds, $u$, $v$, and $w$, and temperature $\theta$. We sequentially split the 50-minute dataset into 44 minutes for training, 3 minutes for validation, and 3 minutes for testing. The training set is used to train both the FNO and Local-FNO models, while the validation set is reserved for hyperparameter tuning and early stopping. The test set is solely used to evaluate model performance, remaining fully inaccessible during training.

Our goal is to use machine learning models to predict multivariable 3D urban microclimate across full grids of $500 \times 500 \times 150$. 
However, training models directly on these full grids would require extensive computational resources that exceed our device capabilities. 
To make training feasible, we utilize FNO’s resolution-invariant property: FNO trained on downsampled data can be applied directly to predict at the original resolution. 
Therefore, we downsample the full grid data from $500 \times 500 \times 150$ to a half-resolution grid of $250 \times 250 \times 75$ for training. 
FNO trains on the entire domain, while Local-FNO further divides the domain into smaller patches for training. 

We train both FNO and Local-FNO using root mean square error (RMSE) loss. Let $ \hat{u} $, $ \hat{v} $, $ \hat{w} $, and $ \hat{\theta} $ be the predicted variables, and $u$, $v$, $w$, and $ \theta $ be the ground truth. Let $ s_x $, $ s_y $, and $ s_z $ represent the spatial resolution of the sample along the $x$, $y$, and $z$ axes, respectively, and let $ N_{\text{batch}} $ denote the number of samples per batch. The RMSE loss is given by:
\begin{equation}
\text{loss}_{\text{RMSE}} = \sqrt{\frac{\sum_{n,x,y,z} \left( (\hat{u}_{n,x,y,z} - u_{n,x,y,z})^2 + (\hat{v}_{n,x,y,z} - v_{n,x,y,z})^2 + (\hat{w}_{n,x,y,z} - w_{n,x,y,z})^2 + (\hat{\theta}_{n,x,y,z} - \theta_{n,x,y,z})^2 \right)}{N_{\text{batch}} s_x s_y s_z} } \ .
\end{equation}
To ensure a fair comparison, both FNO and Local-FNO models with varying patch numbers use the same model architecture: 4 layers, a hidden dimension of 36, and truncation modes of 16, 16, and 8 for the $x$, $y$, and $z$ axes, respectively. 
This setup allows vanilla FNO to train on half-resolution grid data using a 32 GB GPU with a batch size of 1. 
The training hyperparameters are consistent across models, with a learning rate of $2.0\times10^{-3}$, a step scheduler reducing the learning rate by 0.5 per epoch, and the Adam optimizer with a weight decay of $1.0\times10^{-4}$. 
The only differing hyperparameter between FNO and Local-FNO is the batch size. 
Smaller patches require less memory, so we assign a larger batch size to models with smaller patches to fully utilize the 32 GB of GPU memory. 
The batch sizes for each model are shown in Table \ref{tab:batch}. $8 \times 8$ denotes that the entire domain is divided into 64 local patches.
All Local-FNO models use the same overlap rate of 20\%. With patch overlapping, some wind field grid data points are computed multiple times within a single epoch, which increases the training time. A 20\% overlap rate can result in approximately $120\% \times 120\% = 144\%$ of the training time per epoch compared to training without overlap. 
For the implemented models, the training time per epoch is 0.40 hours for FNO and Local-FNO without patch overlapping, and 0.62 hours for Local-FNO with 20\% patch overlapping.

\begin{table}[htbp]
\centering
\begin{tabular}{c|c|c|c|c}
\hline
Model                    & Training patch size    & Batch size & Epochs at stop & Learning rate at stop \\ \hline
FNO                      & $250\times250\times75$ & 1          & 4              & $2.5\times10^{-4}$   \\
Local-FNO ($2\times2$) & $139\times139\times75$   & 3          & 6              & $6.3\times10^{-5}$   \\
Local-FNO ($4\times4$) & $74\times74\times75$     & 10         & 7              & $3.1\times10^{-5}$   \\
Local-FNO ($8\times8$) & $38\times38\times75$     & 42         & 9              & $7.8\times10^{-6}$   \\ \hline
\end{tabular}
\caption{Training specifics for FNO and Local-FNOs with 20\% of patch overlapping and varying patch numbers. The prediction interval is 20 s. All models consume nearly the same amount of GPU memory.}
\label{tab:batch}
\end{table}
To prevent overfitting, particularly as vanilla FNO tends to overfit on the entire domain, we employ an early stopping technique.
Validation loss is monitored each epoch, and training is stopped if the validation loss starts increasing. 
The model with the lowest validation loss is retained. The maximum number of epochs is set to 12.
Table \ref{tab:batch} reports the epochs at stop and learning rates at stop for FNO and Local-FNOs with different patch numbers at a 20 s prediction interval.
Results show that with increased patch numbers and smaller patch sizes, Local-FNO can be trained for more epochs at a lower learning rate without overfitting.
The prediction interval also influences the stopping epochs for all models: with shorter intervals, where flow field changes are minimal, the model learns more effectively and tends to overfit less. 
Conversely, larger intervals with greater flow changes make the model more prone to overfitting.

\section{Results and discussion}

In this section, we evaluate the performance of Local-FNO in predicting multivariable 3D urban microclimate. We begin with an integral time scale analysis to highlight the rapid temporal changes in the flow field. Next, we evaluate Local-FNO’s rollout prediction performance, examining both the predicted instantaneous flow fields and statistical flow characteristics. Finally, we demonstrate the performance improvement of Local-FNO compared upon FNO.

\subsection{Integral time scale analysis of the flow field}
\label{sec:time}

Turbulent systems at different scales exhibit varying levels of complexity. To gain a clearer understanding of the temporal dynamics of the urban microclimate simulation data, we perform an integral time scale analysis. The integral time scale represents the timespan over which fluctuations in turbulent flow remain correlated \cite{batchelor1953theory}. A shorter integral time scale means that turbulent structures lose correlation more quickly, suggesting a more chaotic or rapidly changing flow. 

\begin{figure} [htbp]
    \centering
    \includegraphics[width=0.6\textwidth]{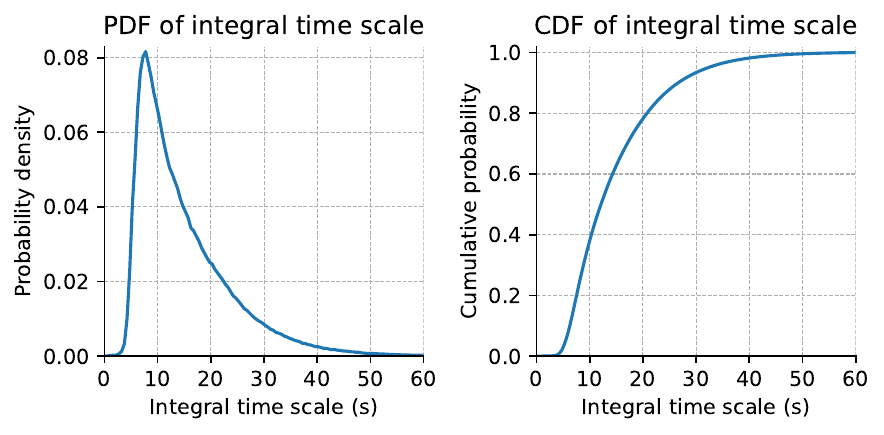}
    \caption{Probability density function (PDF) and cumulative distribution function (CDF) of the integral time scale in the central $1.5 \text{ km} \times 1.5 \text{ km} \times 100 \text{ m}$ domain with compact building distribution.}
    \label{fig:time_curve}
\end{figure}

To compute the integral time scale for the $u$ component of the wind speed time series $u_{t(i)}$, $i = 1, \dots, N$ at a specific location, we first calculate its temporal autocorrelation $R_u(\tau)$ at each time lag $\tau$: 
\begin{equation}
R_u(\tau) = \frac{\sum_{i=1}^{N-\tau} \left( u_{t(i)} - \overline{u} \right) \left( u_{t(i + \tau)} - \overline{u} \right)}{\sum_{i=1}^{N-\tau} \left( u_{t(i)} - \overline{u} \right)^2} \ ,
\end{equation}
where $\overline{u} = \frac{1}{N} \sum_{i=1}^N u_{t(i)}$ is the time averaged $u$ velocity.
We then find the time lag at the first zero crossing, $\tau_{\text{max}}$. The integral time scale for the $u$ component is defined as the area under the autocorrelation curve up to $\tau_{\text{max}}$:
\begin{equation}
\mathcal{T}_u = \sum_{\tau = 0}^{\tau_{\text{max}}} R_u(\tau) \Delta t \ .
\end{equation}
The final integral time scale is reported as an average of the $u$, $v$, and $w$ components:
\begin{equation}
\overline{\mathcal{T}} = \frac{\mathcal{T}_u + \mathcal{T}_v + \mathcal{T}_w}{3} \ .
\end{equation}
We compute the integral time scale across all locations within in a $1.5 \text{ km} \times 1.5 \text{ km} \times 100 \text{ m}$ domain with compact building distribution for 10 minutes of period and present the statistics in Figure \ref{fig:time_curve}.

Figure \ref{fig:time_curve} shows that approximately 50\% of locations have an integral time scale under 12 s, and around 80\% have an integral time scale under 20 s. The PDF peaks at 8 s, indicating the most frequent time scale. The average integral time scale is 14.9 s, indicating that velocity fluctuations in the flow field lose correlation roughly every 15 s, making accurate turbulence prediction a challenging task.

\subsection{Evaluation of perdiction performance of Local-FNO}

In this section, we assess the prediction performance of Local-FNO on full grid microclimate data ($500\times500\times150$) with three directional velocity components $u$, $v$, $w$, and the temperature $\theta$.

\subsubsection{Rollout prediction performance}

\begin{figure} [htbp]
    \centering
    \includegraphics[width=1\textwidth]{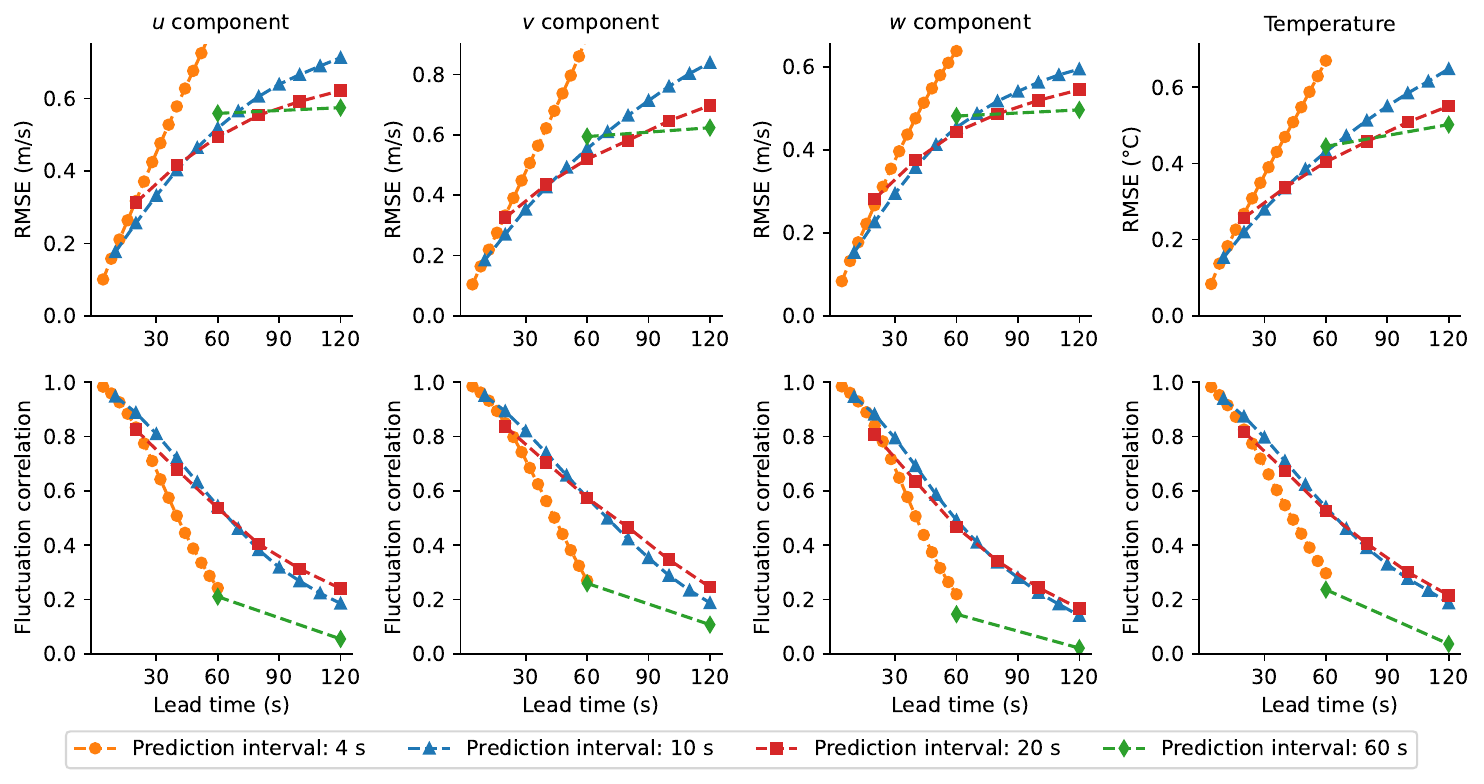}
    \caption{Rollout prediction performance of Local-FNOs with varying prediction intervals.}
    %evaluated using RMSE and fluctuation correlation.
    \label{fig:interval_curve}
\end{figure}

 Unlike CFD, where CFL conditions restrict time steps \cite{cfl}, machine learning methods allow greater flexibility in setting the prediction interval. Shorter prediction intervals improve immediate accuracy but accumulate more errors over time. Conversely, larger prediction intervals reduce the frequency of error accumulation but may miss rapid weather changes, leading to blurrier results. Figure \ref{fig:interval_curve} illustrates the impact of prediction interval length on Local-FNO’s performance. Local-FNO is configured with $8 \times 8$ patches and a 20\% overlap, as this setup yields relatively optimal performance (see section \ref{sec:improvements} for details).

We use two metrics to evaluate prediction performance: RMSE and fluctuation correlation. 
RMSE measures the average error magnitude between predicted results and simulation data, with lower values indicating that predictions are generally closer to observed values. 
The RMSE for a variable $u$ is defined as \begin{equation}
\text{RMSE}_u = \sqrt{\frac{1}{s_x s_y s_z} \sum_{x, y, z} \left( \hat{u}_{x,y,z} - u_{x,y,z} \right)^2} \ ,
\end{equation} 
where $\hat{u}$ and $u$ represent the prediction and ground truth, respectively, and $s_x$, $s_y$, and $s_z$ are the spatial resolutions along the $x$, $y$, and $z$ axes. 
The computation is similar for other variables.
RMSE alone, however, only captures the average error magnitude and does not reflect changes in turbulence patterns over space and time. 
To evaluate how well the prediction captures wind field fluctuations, we also report the fluctuation correlation. 
The fluctuation correlation $R_{\text{fluc}, u}$ for a variable $u$ is computed as 
\begin{equation}
R_{\text{fluc}, u} = \frac{\sum_{x, y, z}(\hat{u}_{x,y,z} - \bar{u}_{x,y,z})(u_{x,y,z} - \bar{u}_{x,y,z})}{\sqrt{\sum_{x, y, z}(\hat{u}_{x,y,z} - \bar{u}_{x,y,z})^2} \sqrt{\sum_{x, y, z}(u_{x,y,z} - \bar{u}_{x,y,z})^2}} \ ,
\end{equation}
where $\bar{u}$ is the steady-state time-averaged value of $u$ for the simulation data. $\bar{u}$ remains stable after the wind field has developed.
A higher fluctuation correlation indicates that the prediction accurately represents the spatial and temporal dynamics of the wind field. 
A model that only captures time-averaged flow may still have a low RMSE but will show a fluctuation correlation close to 0. 
In weather forecasting, fluctuation correlation is often referred to as the anomaly correlation coefficient (ACC) and is commonly used to evaluate the prediction performance \cite{pangu, neuralgcm}.

As shown in Figure \ref{fig:interval_curve}, RMSE values increase with longer lead times for all variables, indicating that prediction error grows over time. 
Fluctuation correlation values also decrease for all variables as lead time extends, reflecting a decline in accuracy for capturing turbulence patterns. 
A shorter prediction interval, such as 4 s, has the lowest initial RMSE but accumulates errors quickly. 
A longer interval, like 60 s, shows slower RMSE growth but suffers from low fluctuation correlation, missing rapid spatial and temporal fluctuations patterns. 
Moderate intervals like 10 and 20 s provide the best overall performance: for the first 60 s, a 10 s interval achieves the lowest RMSE and highest fluctuation correlation, while after 60 s, a 20 s interval gives the most satisfactory results. 
These intervals, 10 and 20 s, are 50 and 100 times longer than the CFD time step of 0.2 s, demonstrating that machine learning methods can operate effectively with a much longer prediction interval than CFD simulations. 
This advantage is a key reason why machine learning methods can be more efficient than CFD solvers in simulating turbulence.

\begin{table}[htbp]
\centering
\small 
\begin{tabular}{c|c|cccc}
\hline
Prediction period      & Metric (time-averaged)  & $u$ component & $v$ component & $w$ component  & Temperature  \\ \hline
0-60 s             & RMSE                    & 0.358 m/s & 0.380 m/s & 0.316 m/s & 0.300 $^{\circ}\mathrm{C}$ \\ %\cline{2-6}
                       & Fluctuation correlation & 0.757 & 0.772 & 0.732 & 0.747 \\ \hline
0-120 s            & RMSE                    & 0.498 m/s & 0.534 m/s & 0.442 m/s & 0.419 $^{\circ}\mathrm{C}$ \\ %\cline{2-6}
                       & Fluctuation correlation & 0.499 & 0.529 & 0.443 & 0.490 \\ \hline
\end{tabular}
\caption{Time-averaged rollout prediction performance of Local-FNO using the best-performing prediction intervals.}
\label{tab:error}
\end{table}

In Table \ref{tab:error}, we present the prediction performance of Local-FNO with its best-performing prediction interval. 
For the prediction period from 0 to 60 s, RMSE values for the three directional wind speeds and temperature remain low (e.g., 0.380 m/s for the $v$ component and 0.300 $^{\circ}\mathrm{C}$ for temperature). 
Fluctuation correlations are high, with values above 0.732 for all components, indicating that Local-FNO effectively captures dynamic fluctuation patterns in the wind field.
As discussed in Section \ref{sec:time}, the average integral time scale of the wind field is 14.9 s. 
Thus, a 60 s prediction period—four times the integral time scale—demonstrates that Local-FNO can provide accurate predictions even after substantial changes in the flow field. 
Even as the prediction period increases to 120 s, the RMSE values remain relatively low (e.g., 0.534 m/s for the $v$ component and 0.419 $^{\circ}\mathrm{C}$ for temperature).

\subsubsection{Visualization of instantaneous flow field}

\begin{figure}[htbp]
    \centering
    \includegraphics[width=1\textwidth]{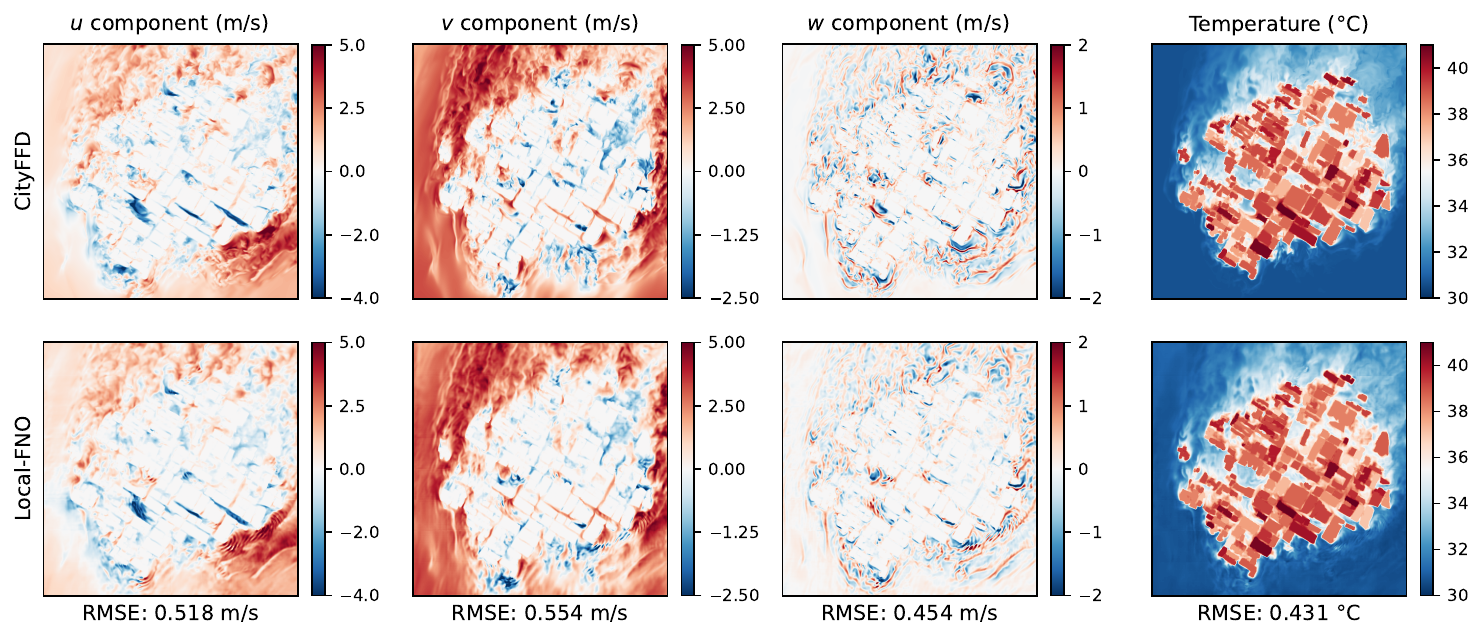}
    \caption{Instantaneous flow field from simulation and Local-FNO prediction, visualized on the $z = 10 \text{ m}$ horizontal plane at 60 s.}
    \label{fig_uvwt_1}
\end{figure}

\begin{figure}[htbp]
    \centering
    \includegraphics[width=1\textwidth]{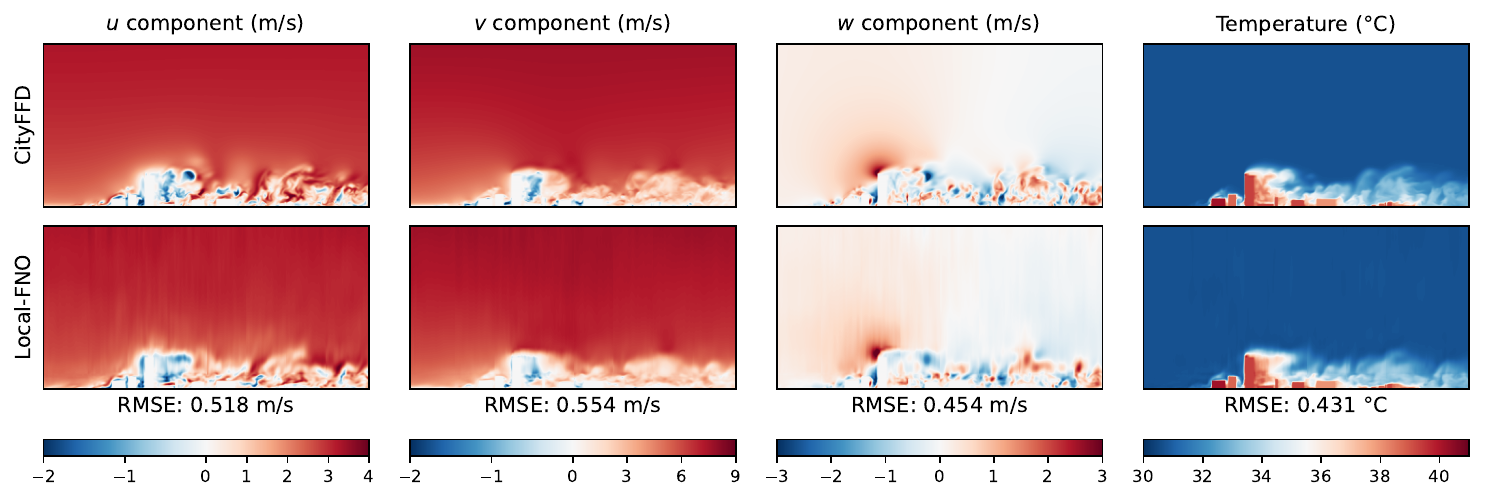}
    \caption{Instantaneous flow field from simulation and Local-FNO prediction, visualized on the vertical plane along the $y$-axis intersecting the tallest building at 60 s.}
    \label{fig_uvwt_2}
\end{figure}

%Tian's 2st version
Figure \ref{fig_uvwt_1} and Figure \ref{fig_uvwt_2} present instantaneous visualizations of the velocity components $u$, $v$, $w$, and the temperature $\theta$ from both CityFFD simulation and Local-FNO prediction at a lead time of 60 s, shown in the horizontal and vertical planes, respectively.
% Figure \ref{fig_uvwt_1} and Figure \ref{fig_uvwt_2} present instantaneous visualizations of the velocity components ($u$, $v$, $w$) and the temperature ($T$) from both the CityFFD calculations and the local FNO predictions in the horizontal and vertical planes, respectively. 
In the first three columns of the velocity fields, CityFFD results highlight pronounced high-speed zones around buildings, along with vortex shedding in their wake. In dense urban environments, airflow passes through narrow gaps and open channels between buildings, creating alternating regions of acceleration and deceleration. 
%In the first three columns of velocity fields, the CityFFD results highlight pronounced high-speed zones around buildings, along with vortex shedding in their wake. In densely packed urban environments, airflow passes through narrow gaps and open channels between buildings, creating alternating regions of acceleration and deceleration. 
Flow separation at building edges downstream generates eddies, inducing vortices that subsequently detach from the structures and create meandering flow patterns in wake regions (see the first two columns in Figure \ref{fig_uvwt_1}). These vortices interact with building roofs, contributing to complex recirculation zones and the formation of mixing layers within the urban canopy layer (see Figure \ref{fig_uvwt_2}).
%Downstream of these building areas, eddy forms due to flow separation at the building edges. This separation induces vortices that subsequently detach from the structures, generating meandering flow patterns in the wake regions (see first two columns in Figure \ref{fig_uvwt_1}). These vortices interact with the roofs of buildings, further contributing to complex recirculation zones and the formation of mixing layers within the urban canopy layer (see Figure \ref{fig_uvwt_2}).
Local-FNO effectively captures primary structural characteristics of urban airflow, such as flow separation, vortex shedding, and turbulent wake effects within urban building environments, achieving RMSE values of 0.518 m/s, 0.554 m/s, and 0.454 m/s across the three directional velocity components. However, minor discrepancies in velocity magnitude remain in highly turbulent regions. 
%The local FNO model effectively captures the primary structural characteristics of urban airflow, including flow separation, vortex shedding, and turbulent wake effects within urban building environments. However, minor differences in the velocity magnitude persist in regions with high turbulence, with the RMSE values of 0.518 $m / s$, 0.554 $m / s$, and 0.454 $m / s$ across the three velocity components. Therefore, further refining the training procedures around zones with high gradient variations may enhance predictive performance in these challenging regions.

In the temperature field, hotter regions in the CityFFD simulation results are primarily located within dense building areas (see Figure \ref{fig_uvwt_1}). In these zones, stable airflow restricts the mixing of warm and cool air, causing heat to build up.  Additionally, the close proximity of buildings reduces natural ventilation, further restricting heat dispersion (see Figure \ref{fig_uvwt_2}). In contrast, cooler areas appear in open spaces and along wider pathways, where freer airflow allows for more efficient heat dissipation. 
% Furthermore, hotter areas shown in the CityFFD simulation results are located mainly inside the densely packed building areas (see Figure \ref{fig_uvwt_1}). In these zones, stable airflow restricts the mixing of warm and cool air, causing heat to build up. In addition, the close proximity of buildings reduces natural ventilation, making it harder for heat to disperse (see Figure \ref{fig_uvwt_2}). In contrast, cooler areas are found in open spaces and along wider paths, where freer airflow allows heat to dissipate more effectively. 
The Local FNO model accurately captures these temperature patterns, including thermal gradients and heat build-up around buildings, achieving an RMSE of 0.431 $^{\circ}\mathrm{C}$. Minor discrepancies arise near building edges and wake regions, where complex temperature dynamics arise due to flow separation and vortex shedding.
% The local FNO model accurately captures these patterns of temperature variation, including thermal gradients and heat build-up around buildings, achieving an RMSE of $0.431 {^\circ C}$. Minor discrepancies appear near the building edges and wake regions, where complex temperature dynamics arise due to flow separation and vortex shedding. Consequently, further refinement may be needed to accurately represent extreme localized thermal effects in areas with high turbulence.

\subsubsection{Visualization of statistical characteristics of flow field}

\begin{figure}[htbp]
    \centering
    \includegraphics[width=0.93\textwidth]{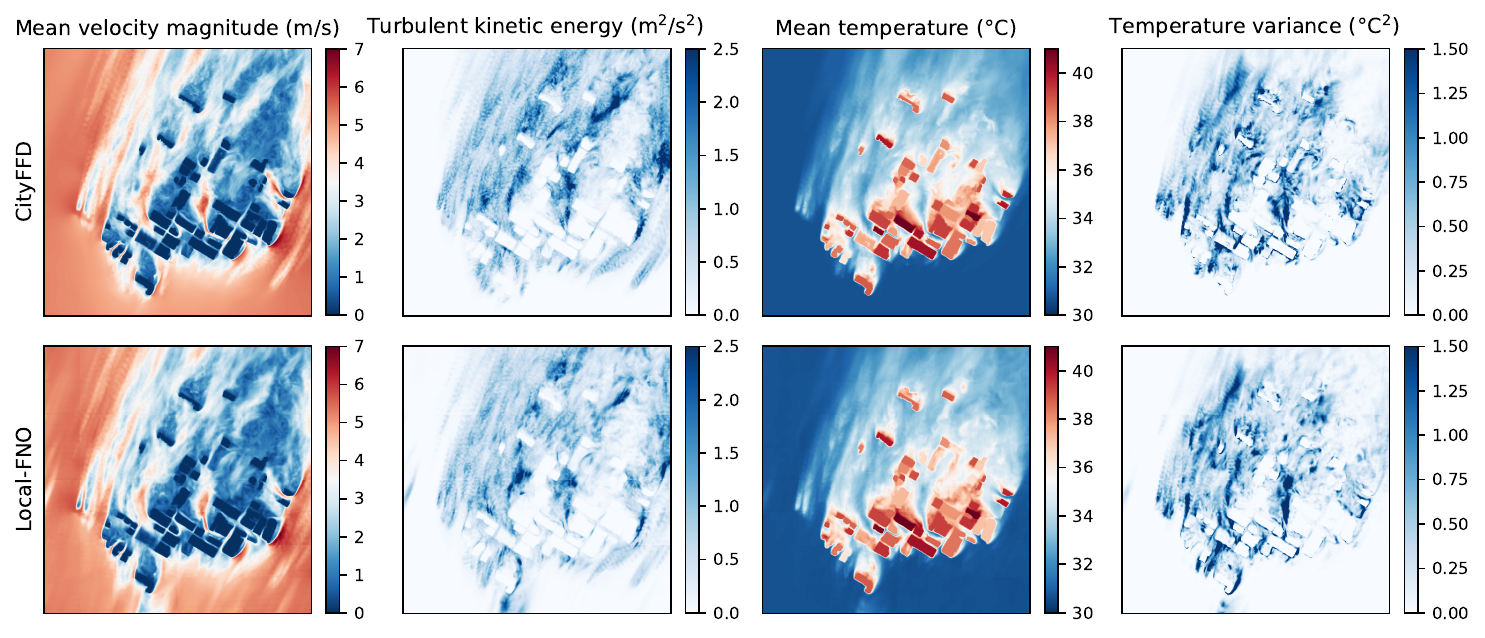}
    \caption{First and second-order flow field statistics from simulation and Local-FNO prediction, visualized on the $z = 50 \text{ m}$ horizontal plane over the 0-120 s period.}
    \label{fig_mean}
\end{figure}

In addition to the instantaneous flow fields, we further examine the statistical characteristics of Local-FNO predictions over a time period. Figure \ref{fig_mean} presents horizontal slices of the mean velocity magnitude, turbulent kinetic energy, mean temperature, and temperature variance from both CityFFD simulations and Local-FNO predictions over the 0-120 s period, with a 50 m height selected to highlight complex urban turbulent flow.
% Figure \ref{fig_mean} presents horizontal slices of the mean velocity magnitude, turbulent kinetic energy, mean temperature, and temperature variances, allowing a detailed comparison between the CityFFD (top row) and Local-FNO prediction (bottom row). 

Given $N$ discrete time steps, the mean velocity magnitude for a specific location is computed as:
\begin{equation}
\overline{|U|} = \frac{1}{N} \sum_{i=1}^N \sqrt{u_{t(i)}^2 + v_{t(i)}^2 + w_{t(i)}^2} \ .
\end{equation} 
The turbulent kinetic energy (TKE) is defined as:
\begin{equation}
\text{TKE} = \frac{1}{2} \frac{1}{N} \sum_{i=1}^N  \left( \left( u_{t(i)} - \overline{u}\right)^2 + \left( v_{t(i)} - \overline{v}\right)^2 + \left( w_{t(i)} - \overline{w}\right)^2  \right) \ ,
\end{equation} 
where $\overline{u} = \frac{1}{N} \sum_{i=1}^N u_{t(i)}$, and similarly for $\overline{v}$ and $\overline{w}$.

The mean temperature is given by $\overline{\theta} = \frac{1}{N} \sum_{i=1}^N \theta_{t(i)}$, and the temperature variance is computed as:
\begin{equation}
\sigma_\theta^2 = \frac{1}{N} \sum_{i=1}^N \left( \theta_{t(i)} - \overline{\theta}\right)^2 \ .
\end{equation} 

% Tian 2st version

In the CityFFD results, high-velocity zones were observed within narrow gaps and along building edges, accompanied by high turbulence intensity (first two columns). These rapid directional changes create shear layers, where differences in velocity induce instability. This instability further causes the rolling and rotation of eddies, which promotes turbulence generation and increases TKE in the flow field.
% In the CityFFD results, high-velocity zones were observed within narrow gaps and along the edges of buildings, accompanied by high turbulence intensity (first two columns). These abrupt directional changes lead to the formation of shear layers, where velocity differences induce instability. This instability further causes the rolling and rotation of eddies, which promotes turbulence generation and increases the turbulent kinetic energy (TKE) in the flow field. Notably, high-speed airflow around the building leaves a sheltered area that forms a low-pressure wake zone characterized by low-speed airflow.
In terms of temperature, the mean temperature plot (third column) shows elevated temperatures concentrated around areas with dense buildings, where restricted airflow leads to thermal buildup. The temperature variance plot (fourth column) highlights areas with variable thermal conditions, indicating strong temperature mixing. Higher temperature variance is apparent near building edges and in wake regions behind buildings, where turbulence enhances convective mixing and causes intermittent interactions between cooler air and warmer surfaces.
% Moreover, elevated temperatures are concentrated around areas with densely packed buildings, as observed in the mean temperature plots (third column). These regions experience significant thermal buildup due to restricted airflow, which limits heat dissipation and causes temperatures to peak. The red zones mark these high-temperature areas, underscoring zones where heat is trapped due to minimal ventilation and constrained airflow patterns. The temperature variance (fourth column) highlights regions with variable thermal conditions, signaling areas of strong temperature mixing. Higher temperature variance is observed near the edges of buildings and in wake regions behind buildings, where airflow becomes more turbulent. These turbulent zones facilitate convective mixing, creating patches where cooler air intermittently interacts with warmer surfaces. The high variance values, represented in darker blue shades, demonstrate how airflow disturbances around buildings contribute to a complex and uneven thermal environment.

Local-FNO’s predictions are highly consistent with CityFFD for first-order statistical characteristics, such as mean velocity magnitude and mean temperature. 
Local-FNO effectively captures steady-state spatial patterns of both velocity and temperature. 
For velocity, Local-FNO replicates high-velocity regions formed by street canyons and low-velocity regions due to building drag effects, similar to CityFFD. 
For temperature, it accurately captures warmer zones near buildings and the temperature gradients between buildings and open spaces. 
For second-order characteristics, such as TKE and temperature variance, Local-FNO generally aligns well with CityFFD, particularly around building edges and within street canyons where fluctuations are largest. 
However, at certain highly turbulent locations, Local-FNO shows slight intensity differences. This is likely due to the complex high-frequency interactions that these second-order statistics represent, which are challenging to capture precisely. 
Overall, Local-FNO achieves notable accuracy in replicating CityFFD’s airflow and thermal distributions, underscoring its reliability in modeling complex urban wind dynamics.
%The comparison between the CityFFD and Local-FNO models shows a strong alignment in capturing the largest value in densely packed areas and high-variance zones around building edges across mean velocity, turbulent kinetic energy (TKE), and temperature fields. Both methods display similar mean velocity patterns, with reduced velocities in densely built areas and increased velocities in open spaces. The TKE patterns also align well with turbulent structures concentrated near building edges. However, local-FNO shows slight intensity variations, likely due to limitations in capturing high-frequency features. Overall, Local-FNO demonstrates notable accuracy in replicating CityFFD’s thermal and airflow distributions, highlighting its reliability in modeling complex urban dynamics.

\begin{figure}[htbp]
    \centering
    \includegraphics[width=0.8\textwidth]{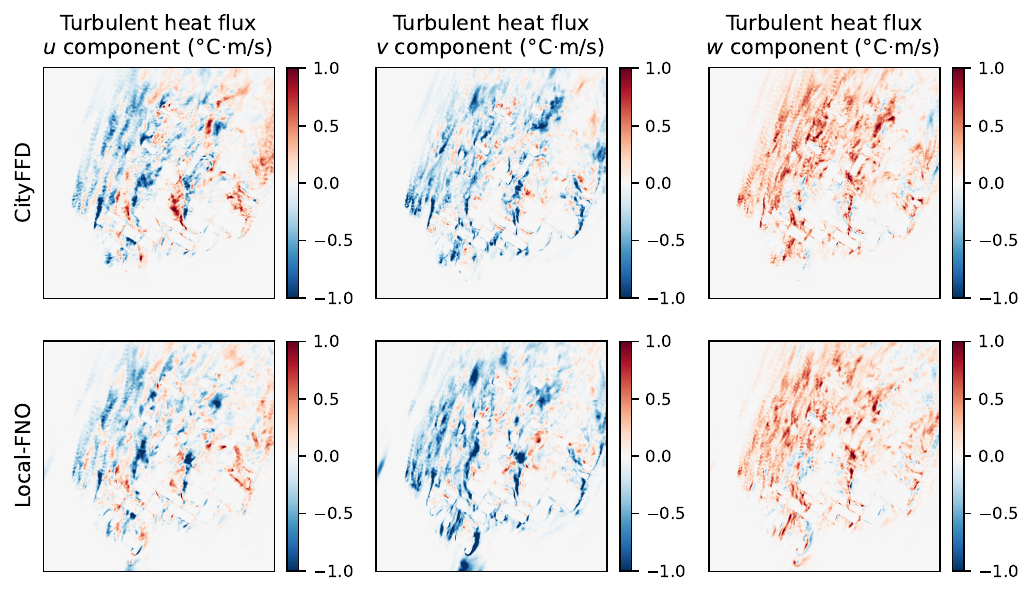}
    \caption{Turbulent heat flux in three directions from simulation and Local-FNO prediction, visualized on the $z = 50 \text{ m}$ horizontal plane over the 0-120 s period.}
    \label{fig_heat_flux}
\end{figure}

% Tian 2st version
We further analyze the statistical coupling between velocity and temperature. Figure \ref{fig_heat_flux} illustrates the turbulent heat flux components ($u$, $v$, and $w$) for both the CityFFD (top row) and Local-FNO (bottom row), with each column corresponding to a different directional heat flux component. For example, the heat flux in the $x$ direction represents the average correlation between temperature fluctuations and $u$ fluctuations, defined as: 
% Figure \ref{fig_heat_flux} illustrates the turbulent heat flux components ($u$, $v$, and $w$) in both the CityFFD (top row) and Local-FNO (bottom row) models, with each column representing a different directional component of heat flux. For example: the heat flux component in the $x$ direction, which is the averaged correlation between temperature fluctuation and $u$ fluctuation, and defined as: 
\begin{equation}
q_x = \frac{1}{N} \sum_{i=1}^N \left( u_{t(i)} - \overline{u}\right) \left( \theta_{t(i)} - \overline{\theta}\right) \ .
\end{equation} 
where $\overline{u}$ and $\overline{\theta}$ are time-averaged $u$ velocity and temperature.

For the $u$ and $v$ components of turbulent heat flux, the alternating red-blue distribution represents dynamic, significant heat transfer influenced by turbulent eddies in densely built areas. Positive zones (red) indicate heat moving along the positive axis direction, while negative zones (blue) indicate cooler air flowing along the positive axis direction. This pattern of alternating colors reflects fluctuating temperature gradients, where intense turbulence drives rapid shifts in heat movement from warmer to cooler regions. Such alternating regions are a direct result of turbulent mixing, particularly prominent in high-density building areas due to vortex shedding and recirculation zones \cite{tian_non_isothermal_2022}. As the primary wind direction aligns more closely with the $y$-axis ($v$ component), crosswind effects around buildings have a stronger impact on the $u$ component, especially around narrow streets and building edges, where lateral turbulence enhances heat transfer.
In the $w$ component, which represents vertical heat flux between the surface and the atmosphere, the plots show predominantly red regions, indicating strong upward heat transfer. These red zones reflect areas where heat moves vertically, facilitating thermal energy exchange between the surface and atmosphere. This upward flux is particularly prominent in wake regions and above structures, where building-induced turbulence and surface heating cause warm air to rise.
%In the $w$ component (third column), representing vertical heat flux between the surface and the atmosphere, the plots show a predominance of red regions, indicating significant upward heat transfer. The red zones reflect areas where heat is transported vertically, contributing to the exchange of thermal energy between the surface and the atmosphere. This upward movement is particularly prominent in wake regions and above structures where warm air rises due to building-induced turbulence and surface heating. Notably, the local-FNO model presents slightly more pronounced red regions compared to CityFFD, suggesting an enhanced sensitivity to vertical eddy structures. This greater extension of red areas might indicate that the Local-FNO model captures more intricate details of vertical heat dissipation, likely due to a higher responsiveness to turbulence and convective effects. However, this could also mean a slight overestimation of upward heat flux in certain regions, as the model may amplify the effects of turbulent vertical mixing.

For all directions, Local-FNO produces a comparable pattern of red and blue regions, indicating that it captures the primary direction and magnitude of heat transport driven by turbulence. 
Some finer details appear slightly smoothed, particularly near building edges and within narrow streets where sharp flux changes are expected. 
In general, Local-FNO provides a strong representation of heat exchange processes in urban microclimate with temperature variations. 
Accurate prediction of this complex heat flux pattern is vital in urban environments as it impacts both local thermal comfort and larger-scale phenomena like urban heat islands.
%It is worth noting that in urban contexts, this complex heat flux pattern is significant as it affects both localized thermal comfort and broader environmental factors, such as urban heat islands. The alternating heat flux indicates how turbulent eddies distribute heat across densely packed areas, resulting in pockets of temperature variation that can impact air quality and heat transfer. Thus, the local FNO's detailed representation of these turbulent structures highlights its ability to simulate complex thermal dynamics in urban environments.

\subsubsection{Improvement from patch overlapping}

\begin{figure} [htbp]
    \centering
    \includegraphics[width=0.5\textwidth]{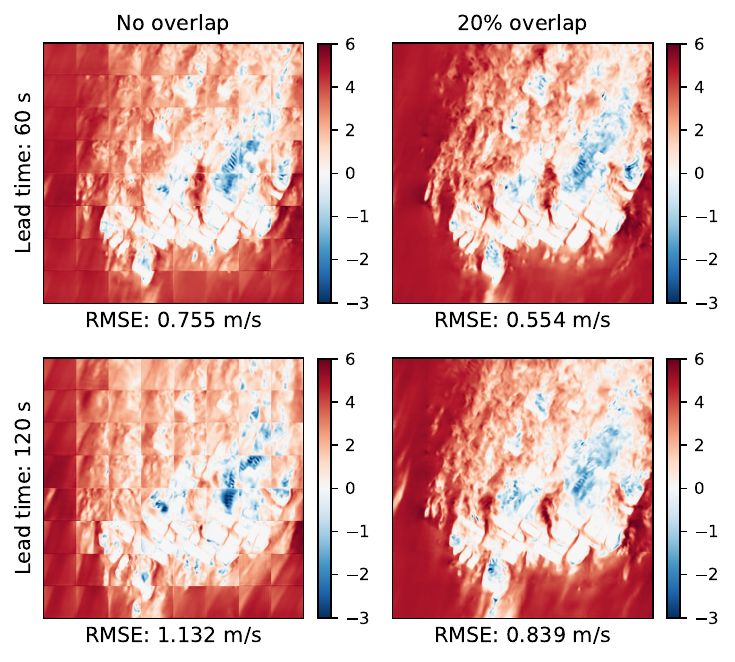}
    \caption{Instantaneous predictions of the $v$ component from Local-FNOs without and with patch overlapping, visualized on the $z = 50 \text{ m}$ horizontal plane at 60 s and 120 s.}
    \label{fig:overlap}
\end{figure}

Patch overlapping is a key component in Local-FNO to ensure smoother transitions between patches. 
Figure \ref{fig:overlap} shows a comparison of Local-FNO ($8\times8$) predictions without and with patch overlapping. 
Without overlap, Local-FNO's prediction exhibits grid artifacts in blocky patterns, which become more pronounced as the prediction lead time increases from 60 s to 120 s. 
These artifacts arise because the air in the actual flow field continuously interacts across neighboring regions, transferring both kinetic energy and heat. 
As lead time increases, Local-FNO's independent local predictions without overlap make patches drift further from the actual flow field. 
In contrast, with a 20\% overlap, grid artifacts are significantly reduced, creating smoother transitions across patches and a more coherent spatial structure. 
Even at a lead time of 120 s, Local-FNO with overlapping patches exhibits minimal grid artifacts, demonstrating that patch overlapping effectively facilitates flow interactions across neighboring patches and mitigates the side effects of independent local predictions.

\subsubsection{Computational efficiency}

CityFFD and Local-FNO both leverage GPU acceleration. CityFFD requires 4.6 s to compute per time step, whereas Local-FNO takes 9.8 s for a full grid ($500\times500\times150$) prediction. Therefore, the speedup of Local-FNO depends mainly on its prediction interval. With a CFD time step of 0.2 s and a 20 s prediction interval, Local-FNO achieves a 47 times speedup on full grid microclimate data. As Local-FNO is resolution-invariant, further acceleration can be achieved by predicting on half-resolution grid ($250\times250\times75$). This further reduces the computation time for prediction to 0.95 s and results in a 480 times speedup.

\subsection{Prediction performance improvements of Local-FNO over FNO}
\label{sec:improvements}

This section presents the performance improvements achieved by Local-FNO over the vanilla FNO model. Due to GPU memory constraints, predicting on the full grid microclimate data ($500\times500\times150$) with vanilla FNO exceeds our device’s capacity. Therefore, all models in this section predict on half-resolution grid ($250\times250\times75$) to ensure a fair comparison.

\subsubsection{Improvement in rollout prediction performance}

\begin{figure} [htbp]
    \centering
    \includegraphics[width=1\textwidth]{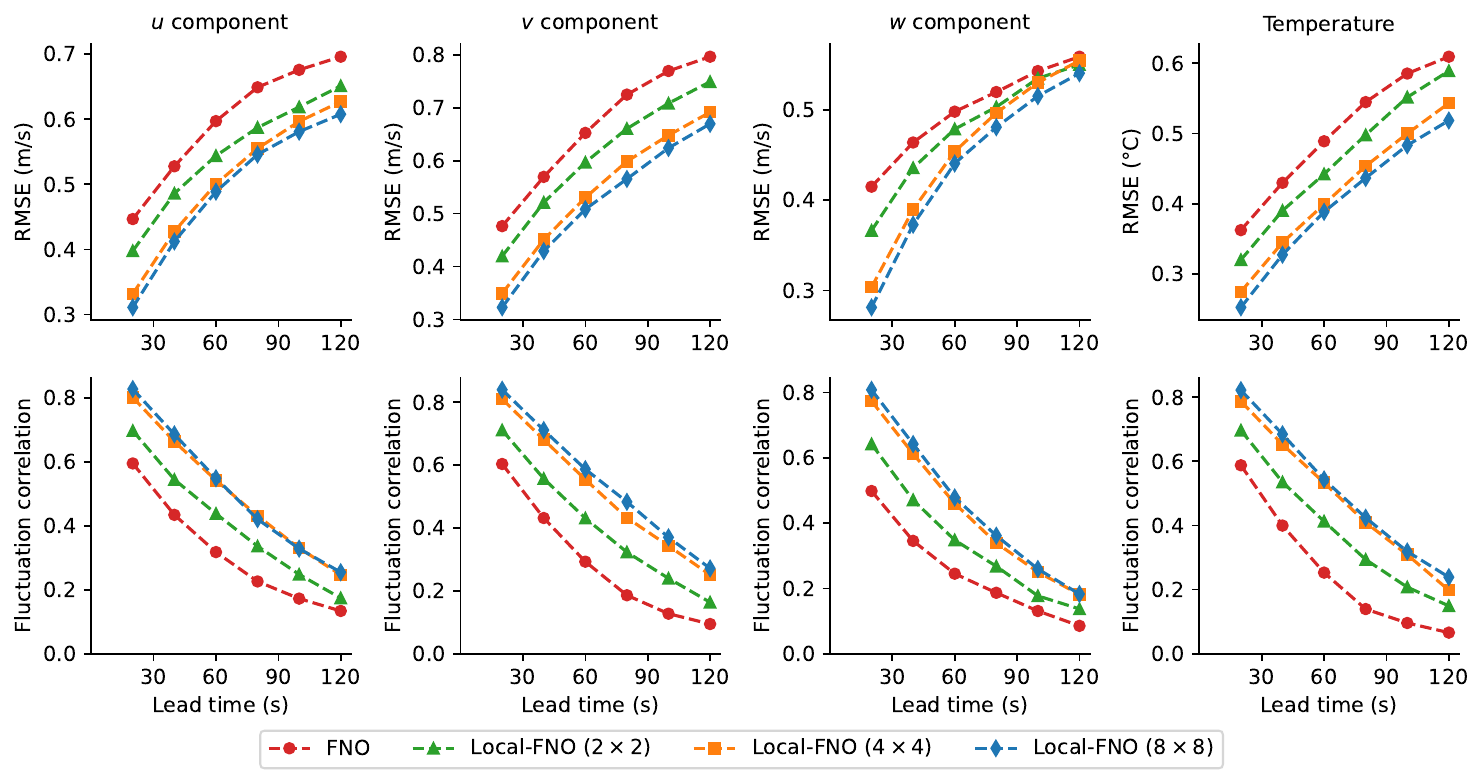}
    \caption{Rollout prediction performance of FNO and Local-FNOs with varying patch numbers.}
    \label{fig:patch_curve}
\end{figure}

Figure \ref{fig:patch_curve} shows the rollout prediction performance comparison between FNO and Local-FNOs with different patch numbers. 
In this context, $8 \times 8$ denotes that the entire domain is divided into 64 local patches. FNO can be treated as a special case of Local-FNO with a single patch. 
As illustrated in Figure \ref{fig:patch_curve}, Local-FNOs outperform FNO across all metrics and lead times, exhibiting lower RMSE and higher fluctuation correlation. 
Performance improvements are particularly notable when increasing from 1 to $2 \times 2$ patches and from $2 \times 2$ to $4 \times 4$ patches. 
However, improvements are limited from $4 \times 4$ to $8 \times 8$ patches. 
This diminishing return trend suggests that increasing the number of patches beyond a certain point yields smaller benefits, with $8 \times 8$ patches providing the best overall performance.
During the 0-60 s period, Local-FNO ($8 \times 8$) shows a significant improvement over FNO in prediction performance, reducing RMSE by 23.5\%, 26.3\%, 21.1\%, and 24.9\% for the $u$, $v$, $w$, and $\theta$ components, respectively, reflecting a lower average error magnitude. Additionally, Local-FNO ($8 \times 8$) achieves fluctuation correlations that are 45.2\%, 50.1\%, 46.0\%, and 47.7\% closer to 1 for the $u$, $v$, $w$, and $\theta$ components, reflecting greater accuracy in capturing turbulence pattern fluctuations over space and time.
% average: 23.9% and 47.3%

\subsubsection{Improvement in instantaneous flow field}

\begin{figure} [htbp]
    \centering
    \includegraphics[width=1\textwidth]{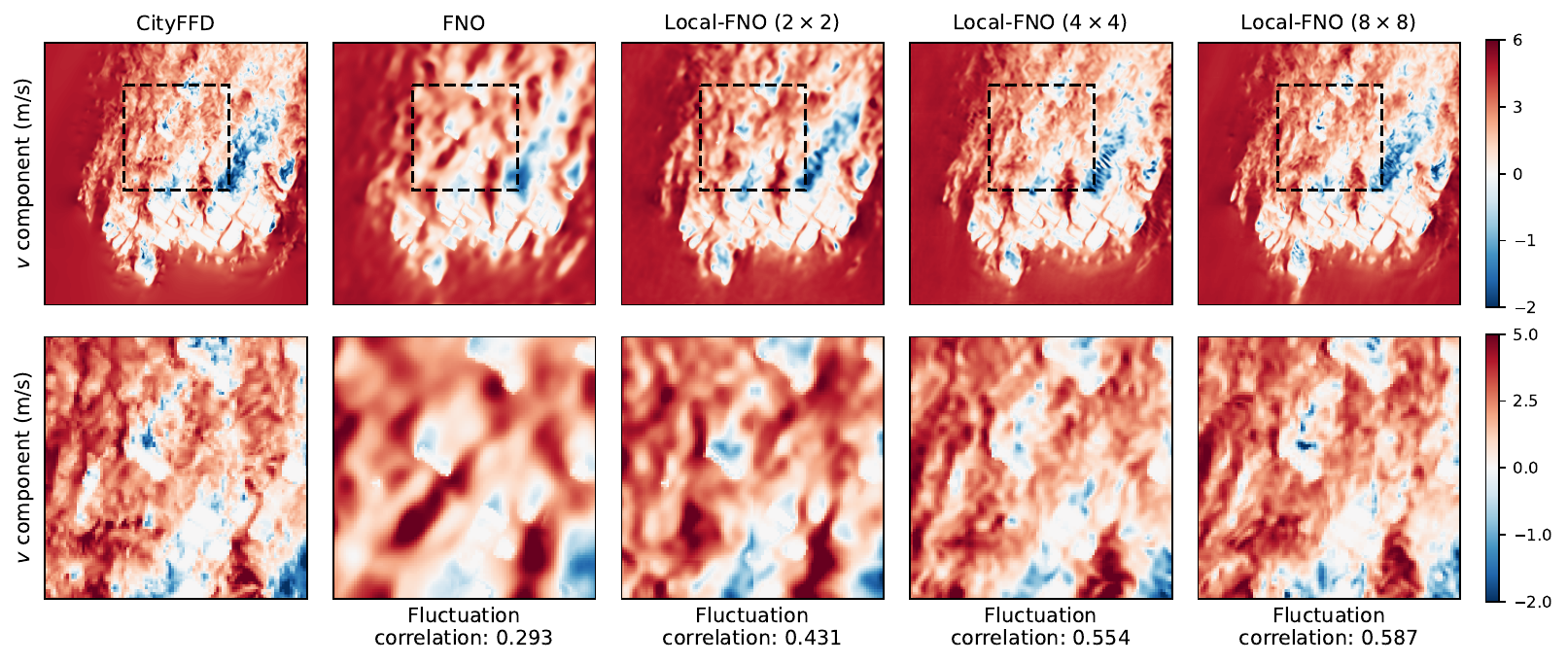}
    \caption{Instantaneous predictions of the $v$ component from FNO and Local-FNOs with varying patch numbers, visualized on the $z = 50 \text{ m}$ horizontal plane at 60 s.}
    \label{fig:patch_v}
\end{figure}

\begin{figure} [htbp]
    \centering
    \includegraphics[width=1\textwidth]{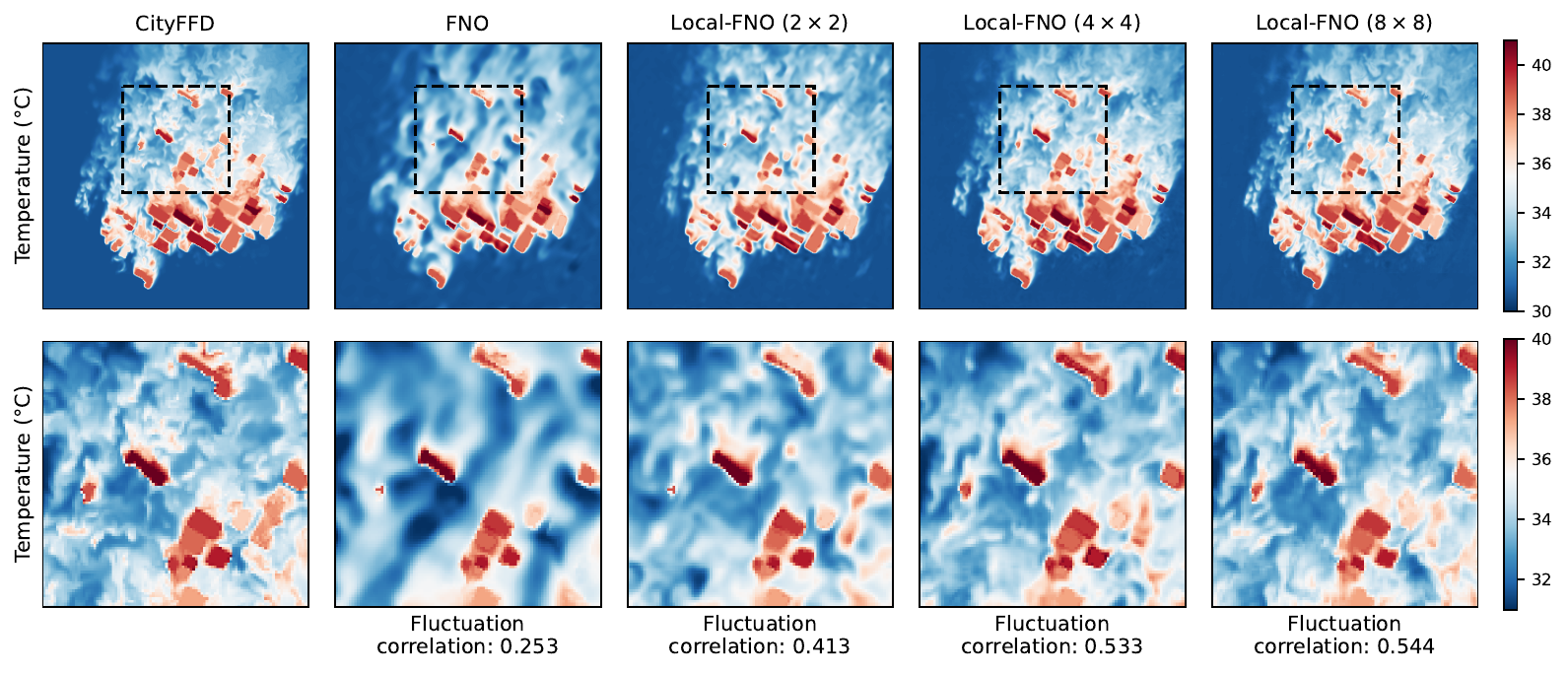}
    \caption{Instantaneous temperature predictions from FNO and Local-FNOs with varying patch numbers, visualized on the $z = 50 \text{ m}$ horizontal plane at 60 s.}
    \label{fig:patch_t}
\end{figure}

Figures \ref{fig:patch_v} and \ref{fig:patch_t} display the instantaneous velocity component $v$ and temperature $\theta$ at a 60 s lead time, comparing predictions from FNO and Local-FNOs with different patch numbers. 
Within 60 s, small-scale turbulence patterns change significantly. 
Compared to the CityFFD simulation, FNO’s predictions appear blurry and fail to capture the small-scale turbulence in both velocity and temperature. 
As discussed in Section \ref{sec:local-fno}, this limitation arises because FNO preserves only a limited number of low-frequency Fourier modes. 
Fourier modes defined on larger domains have longer wavelengths, which are ineffective for representing small-scale features. 
Additionally, FNO’s direct training on whole-domain, high-dimensional data makes it prone to overfitting, especially with limited training samples. 
Prior research showed that FNO can deliver satisfactory predictions in simpler wind fields over shorter prediction intervals (0.1 s) \cite{peng2024fourier}. 
However, as turbulence complexity and the prediction interval grow, its predictions degrade, producing a blurry flow field that no longer accurately reflects turbulence characteristics.

Compared to FNO, Local-FNOs provide significantly clearer flow fields for both velocity and temperature. 
As the number of patches increases, the size of each patch decreases, allowing Local-FNO to produce sharper and more accurate flow field predictions. 
This suggests that Fourier modes defined on smaller patches more effectively capture small-scale features in complex turbulence. 
For instance, Local-FNO ($8 \times 8$) improves turbulent fluctuation correlation from 0.293 to 0.587 for the $v$ component and from 0.253 to 0.544 for temperature, compared to FNO.

\subsubsection{Improvement in energy spectra}

\begin{figure} [htbp]
    \centering
    \includegraphics[width=0.4\textwidth]{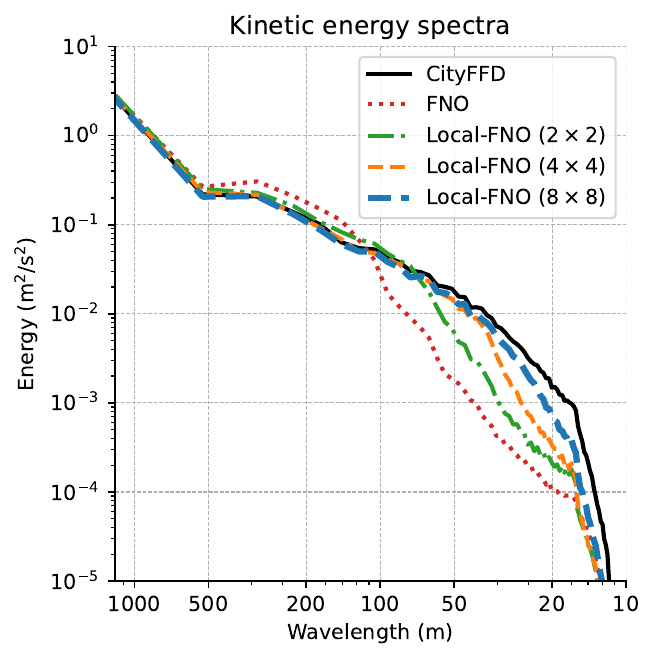}
    \caption{Kinetic energy spectra of velocity field predictions from FNO and Local-FNOs with varying patch numbers, computed on the $z = 50 \text{ m}$ horizontal plane at 60 s.}
    \label{fig:spectra}
\end{figure}

Energy spectra provide a quantitative measure of the blurriness in flow fields, representing how the kinetic energy of turbulence is distributed across various spatial scales \cite{kolmogorov1991local}. 
It's commonly used to evaluate whether machine learning predictions can capture complex multi-scale turbulence patterns. 
Figure \ref{fig:spectra} shows the energy spectra for velocity field predictions from FNO and Local-FNOs with different patch numbers.
In the CityFFD simulation, energy gradually declines as wavelength decreases, showing a smooth, continuous drop across scales. 
The FNO prediction, in comparison, overestimates energy at scales above 100 m and underestimates it at scales below 100 m, indicating a limitation in predicting small-scale turbulence details. 
This limitation is also observed in other studies as the spectral bias of FNO \cite{liu2024mitigating, qin2024toward}.
In urban environments, structures such as dense buildings and pathways can generate complex flow patterns at scales below 100 m. 
Accurate representation of turbulence at these scales is crucial for modeling wind interactions with urban structures. 
Local-FNO, with increased patch numbers and smaller patch sizes, aligns more closely with CityFFD energy spectra across all scales, with significant improvements between 100 m and 20 m. 
A 20 m wavelength corresponds to a 10 m resolution according to the sampling theorem \cite{shannon1949communication}, demonstrating the effectiveness of Local-FNO in accurately predicting complex urban wind flows.

%%\subsection{Limitations}

\section{Conclusions}
\label{sec:conclustions}
In this study, we developed a novel Local-FNO model to achieve accurate and efficient predictions for multivariable, high-resolution 3D urban microclimate. For the rapidly changing turbulence in urban environments, Local-FNO is capable of making accurate predictions of instantaneous flow field over 60 seconds, four times the average integral time scale. 
Additionally, it accurately captures statistical flow characteristics, including turbulent kinetic energy and turbulent heat flux, over a 120 seconds period. This demonstrates Local-FNO’s ability to model both individual variables (three directional velocities and temperature) and their correlations. 
Running on a single 32 GB GPU, Local-FNO can predict high-resolution urban microclimate data with 150 million ($500 \times 500 \times 150 \times 4$) feature dimensions. 
For the same prediction lead time, it achieves nearly 50 times the speed of a CFD solver, with an average error of 0.35 m/s for velocity and 0.30 $^{\circ}\mathrm{C}$ for temperature over the first 60 seconds. 
These findings demonstrate the potential of machine learning for efficiently and accurately solving complex fluid dynamics in urban environments with temperature variations.
Local-FNO is capable of capturing turbulence patterns down to a 10 m resolution within a $2 \text{ km} \times 2 \text{ km}$ domain.
Although Local-FNO does not yet match the precision of numerical solvers, it offers meaningful results at much faster speeds, which can support practical applications like energy-efficient urban planning and carbon reduction. 
Moreover, studies suggest that with larger models and expanded training data, the accuracy of end-to-end trained machine learning models can continue to improve and surpass that of numerical solvers \cite{pangu, graphcast}.

Local-FNO offers valuable insights in designing more accurate and efficient machine learning models. 
By leveraging local Fourier features, it effectively addresses three challenges often encountered with global Fourier features in high-resolution data: blurry output quality, extensive GPU memory usage, and substantial data demands. 
For a given number of Fourier modes, those defined over smaller local regions have shorter wavelengths, enabling the representation of finer-scale details. 
Local Fourier features also better capture spatially varying, non-stationary high-frequency information, similar to the improvements wavelet analysis brings over traditional Fourier analysis \cite{daubechies1990wavelet}.
By dividing full-domain microclimate prediction into independent subtasks, Local-FNO reduces memory usage and improves scalability, opening the door for large-scale predictions across multiple GPUs. 
Training on local patches allows Local-FNO to make efficient use of limited data, enhancing generalization. 
Furthermore, geometry encoding introduces local geometric information, enabling the model to handle fluid dynamics across different geometric boundaries with greater precision. 
Patch overlapping enables interaction between neighboring regions, reducing discontinuities between patches.

This work has several limitations. 
First, Local-FNO is trained and tested in a single city area with a fixed wind profile. 
Although it can predict wind dynamics across different local building geometries within this area, its performance in other city areas and under varied wind profiles requires further study. 
Second, Local-FNO does not account for global wind field features, which works well for predicting small-scale turbulence in fully developed wind fields. 
However, the local training strategy can be less effective when significant large-scale patterns change in the global wind field. 
Third, patch overlapping in Local-FNO leads to some wind field grid data being trained and predicted multiple times, increasing computational time. 
A 20\% overlap rate, for instance, results in approximately 144\% of the computational time compared to non-overlapping training.

For future work, a promising direction is to integrate the strengths of global and local Fourier features. 
Global Fourier features can capture large-scale, low-frequency patterns, while local Fourier features can capture small-scale, high-frequency patterns. 
This approach could yield a model proficient in capturing both large-scale and small-scale dynamics within complex multiscale flow fields. 
Another avenue worth exploring is the coupling of CFD solvers with machine learning models. 
From a frequency perspective, CFD solvers converge more quickly on local, high-frequency features, while machine learning models converge more quickly on global, low-frequency features. 
Together, they could achieve faster computations without compromising accuracy.
 
% \section*{Declaration of competing interest}

% The authors declare no competing financial or non-financial interests.

% \section*{Data availability}

% Other data that support the findings of this study are available from the corresponding author upon reasonable request.

% \section*{Code availability}

% Codes used to generate the results of this study are available from the authors on reasonable request.

% \section*{Acknowledgements}

% The research is supported by the Environment and Climate Change Canada (ECCC) through the Contribution Agreement Program [\#GCXE23S006], the Natural Sciences and Engineering Research Council (NSERC) of Canada through the Discovery Grants Program [\#RGPIN-2024-06297], the Research Scheme of the Research Grants Council of Hong Kong SAR, China [T22-504/21R], and the SEED project “Creating Electrified and Decarbonized Healthy Urban Microclimate around Building Clusters through Climate-Resilient Solutions” under the Canada First Research Excellence Fund (Volt-Age).

% \bibliographystyle{unsrt}
% \bibliographystyle{abbrv}
\bibliography{main}

\end{document}